\documentclass[a4paper,11pt]{article}
\usepackage{jheppub}
\usepackage{braket}
\usepackage{comment}
\usepackage{appendix}
\usepackage{mathtools}
\usepackage{caption}
\usepackage{subcaption}
\usepackage{bbold}
\usepackage[utf8]{inputenc}
\usepackage{tikz-cd}
\usepackage[T1]{fontenc} 
\usepackage{scalerel}
\usepackage{amsmath}
\usepackage{amsthm}

\usepackage{mathrsfs}
\usepackage{physics}
\usepackage[compat=1.1.0]{tikz-feynman}
\usepackage[T1]{fontenc} 
\usepackage{tensor}
\usepackage[]{graphicx}
\usepackage[export]{adjustbox}
\usepackage{slashed}
\usepackage{stackrel,amssymb}
\usepackage{tikz}
\usetikzlibrary{math}
\usetikzlibrary{calc,intersections}
\usetikzlibrary{arrows.meta,shapes.misc}
\usepackage{mathtools}
\usetikzlibrary{arrows.meta}
\usepackage{comment}
\tikzset{>={Latex[scale=1.1]}}
\usetikzlibrary{arrows.meta,
	bending,
	decorations.markings, decorations.text}

\makeatletter
\newcommand*{\letterdef@}{}
\newcommand*{\letterdef}[3]{%
	\def\letterdef@##1{\expandafter\newcommand\csname #1\endcsname{#2{##1}}}%
	\@tfor\@tempa :=#3\do{\expandafter\letterdef@\expandafter{\@tempa}}}
\makeatother
\letterdef{c#1} {\mathcal}{ABCDEFGHIJKLMNOPQRSTUVWXYZ} 
\letterdef{rm#1}{\mathrm} {dDeimM} 
\date{\today}
\author[a]{Carlos Barredo Mart\'inez,}
\author[b]{Torben Skrzypek}
\affiliation[a]{Abdus Salam Centre for Theoretical Physics,\\
Imperial College London, Prince Consort Road, SW7 2AZ, UK.}
\affiliation[b]{Deutsches Elektronen-Synchrotron DESY,\\
Notkestraße 85, 22607 Hamburg, Germany.}
\emailAdd{carlos.barredo21@imperial.ac.uk}
\emailAdd{torben.skrzypek@desy.de}
\abstract{We consider type IIB string theory on $\mathrm{AdS}_5\times S^5/\mathbb{Z}_{L}$ orbifold spaces with generic $L$. Recent localisation results in the dual 4d $\mathcal{N}=2$ circular quiver gauge theories provide us with strong coupling expansions of certain correlators involving twisted half-BPS operators. To leading order, these results have been matched to an effective theory for massless twisted string states, which can be constructed by resolving the orbifold singularity and considering localised supergravity modes on the resolution cycles. Applying this reasoning to subleading order in strong coupling, we observe that for $L\neq 2,3,4,6$, a naive reduction of the 10d $(\alpha')^3$-correction does not result in the correct coefficients to match the localisation result. We explain this mismatch by the appearance of twisted sector resonances in string amplitudes involving external twisted sector states. We perform the low-energy expansion of a ``twisted'' Virasoro-Shapiro amplitude and recover the expected coefficients, suggesting that the orbifold resolution and the low-energy expansion can not be interchanged directly. Finally, we comment on the long-quiver limit, $L\to\infty$, in the context of the low-energy effective action.}

\begin{document}
\begin{flushright}Imperial$-$TP$-$2025$-$CBM$-$01\\DESY-25-190\end{flushright}
\title{Exploring the twisted sector of $\mathbb{Z}_{L}$ orbifolds: Matching $\alpha'$-corrections to localisation}
\maketitle
\section{Introduction}
In its canonical incarnation, the AdS/CFT duality \cite{Maldacena_1999, Witten:1998qj} relates type IIB string theory on $\mathrm{AdS}_5\cross S^5$ background to 4d $\mathcal{N}=4$ Super-Yang-Mills (SYM) theory with gauge group $\mathrm{SU}(N)$. Recently, $\mathcal{N}=2$ orbifolds of these models have received much attention \cite{Galvagno:2020cgq,Beccaria:2021hvt,Billo:2021rdb,Billo:2022gmq,Skrzypek:2022cgg, Billo:2022fnb,Beccaria:2022ypy,Billo:2022lrv,Beccaria:2023kbl,Skrzypek:2023fkr, Korchemsky:2025eyc, Ferrando:2025qkr,lePlat:2025eod} as they provide simple testing grounds for the AdS/CFT duality that retain much of the structure of the $\mathcal{N}=4$ parent theory (e.g. integrability) but are less constrained by supersymmetry. A salient feature of such theories is that observables which are protected in the $\mathcal{N}=4$ SYM theory (e.g. the three-point functions of certain half-BPS operators) may now pick up a non-trivial dependence on the 't Hooft coupling, $\lambda:=g_{\rm{YM}}^2N$, already in the planar limit. In many cases, the remaining amount of supersymmetry still allows for an explicit computation of these quantities, thus probing non-trivial aspects of the AdS/CFT duality and providing us with valuable insights into the behaviour of type IIB string theory on orbifold backgrounds. In this regard, $\mathcal{N}=2$ orbifolds sit in the sweet spot between calculational control and physical interest.

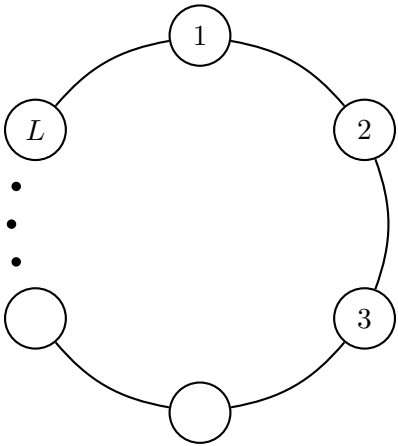
\begin{figure}[htbp]
\centering
\begin{tikzpicture}[
    node/.style={circle, draw, minimum size=0.8cm, thick},
    scale=1.25
]
\node[node] (1) at (90:2) {$1$};
\node[node] (2) at (30:2) {$2$};
\node[node] (3) at (-30:2) {$3$};
\node[node] (b1) at (-90:2) {};
\node[node] (b2) at (-150:2) {};
\node[node] (L) at (150:2) {$L$};

\draw[thick] (L) to[bend left=20] (1);
\draw[thick] (1) to[bend left=20] (2);
\draw[thick] (2) to[bend left=20] (3);
\draw[thick] (3) to[bend left=20] (b1);
\draw[thick] (b1) to[bend left=20] (b2);
\path (b2) to[bend left=20]
    coordinate[pos=0.2] (d1)
    coordinate[pos=0.5] (d2)
    coordinate[pos=0.8] (d3)
(L);
\foreach \p in {d1,d2,d3}
    \fill (\p) circle (0.05);
\end{tikzpicture}
\caption{Diagrammatic representation of the circular quiver theory. Each node represents an $\mathcal{N}=2$ vector multiplet in the adjoint representation of $\mathrm{SU}(N)$. Each line connecting adjacent nodes gives rise to an $\mathcal{N}=2$ hypermultiplet in the bifundamental representation of $\mathrm{SU}(N_{i})\otimes \mathrm{SU}(N_{i+1})$. In the case of $L=2$, an additional $\mathrm{SU}(2)$ symmetry arises among the two hypermultiplets connecting the only two gauge nodes.}
\label{fig:quiverdiagram}
\end{figure}

On the field theory side, the $\mathbb{Z}_L$ orbifold projection can be applied to 4d $\mathcal{N}=4$ SYM with gauge group $\mathrm{SU}(L N)$, resulting in an $\mathcal{N}=2$ superconformal quiver gauge theory with gauge group $\mathrm{SU}(N)^L$, whose matter content is conveniently described by the quiver diagram in Figure \ref{fig:quiverdiagram}. The coupling constants $g_i$ of the various gauge nodes are taken to be equal $g_{\text{YM},i}=g_{\text{YM}}$ \cite{Kachru:1998ys,Lawrence:1998ja,Bershadsky:1998cb,Bershadsky:1998mb,Gukov:1998kk,Klebanov:1998hh,Klebanov:1999rd}.\footnote{From a gauge theory point of view, this is only a one-dimensional subspace of the $L$-dimensional conformal manifold available to these quiver gauge theories. The study of marginal deformations of orbifold theories has been a fruitful tangent \cite{Gadde:2009dj,Gadde:2010zi,Zarembo:2020tpf,Pomoni:2021pbj,Bertle:2024djm,Bozkurt:2024tpz,Bozkurt:2025exl} which goes beyond the scope of this paper.} 

On the string theory side, the dual orbifold backgrounds are constructed by identifying points on the $S^5$ subspace under the $\mathbb{Z}_L$ action
\begin{equation}\label{eq:orbi}
    \Gamma_L:\qquad (X,Y,Z)\quad \to\quad (e^{\frac{2\pi i}{L}}X, e^{-\frac{2\pi i}{L}}Y, Z)\,,
\end{equation}
presented in terms of the flat embedding space $\mathbb{C}^3\supset S^5$. The set of fixed points under this $\mathbb{Z}_L$ action is the six-dimensional submanifold $\mathrm{AdS}_5\cross S^1$, and the local geometry around it takes the form $\mathrm{AdS}_5\cross S^1\cross \mathbb{C}^2/\mathbb{Z}_L$. One may think of the full $\mathrm{AdS}_5\cross S^5/\mathbb{Z}_L$ geometry as the near-horizon limit of a stack of D3 branes located at the singularity of the locally flat orbifold space $\mathbb{R}^{1,5}\cross \mathbb{C}^2/\mathbb{Z}_L$. The spectrum of string theory on such orbifold backgrounds (which we review in Appendix \ref{App: Spectrum}) comprises an untwisted sector of states, which consists of $\Gamma_L$-invariant states of the parent theory, and $(L-1)$ twisted sectors generated by strings which close only up to an action of $\Gamma_L^n$, where $n\in\{1,\dots, L-1\}$. While the dynamics of untwisted states are inherited from the parent theory, the twisted states may behave rather differently. An immediate consequence of their construction is that they can only freely propagate within the fixed $\mathrm{AdS}_5\cross S^1$ subspace. The effective theory that governs the massless twisted sector states at low energies is thus necessarily a six-dimensional theory, in stark contrast to the untwisted sector. This local EFT on the singularity of $\mathbb{R}^{1,5}\cross \mathbb{C}^2/\mathbb{Z}_L$ was identified in \cite{Douglas:1996sw} as an $\mathcal{N}=(2,0)$ supergravity theory, to which the untwisted sector contributes a supergravity and two tensor multiplets and each twisted sector contributes another tensor for a total of $(L +1)$ tensor multiplets. In full $\mathrm{AdS}_5\cross S^5/\mathbb{Z}_L$ background, we expect corrections to this theory.

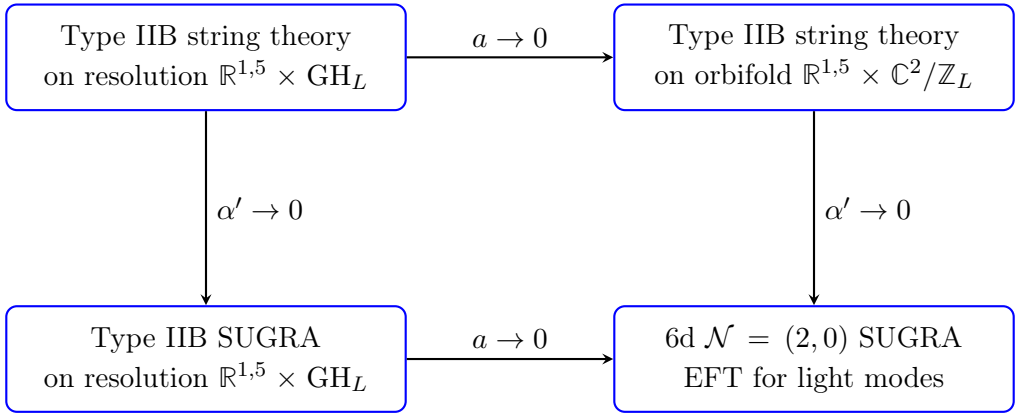
\begin{figure}[htbp]
\centering
\begin{tikzpicture}[
  >=stealth,
  thick,
  box/.style={
    draw=blue,
    rounded corners,
    rectangle,
    align=center,
    text width=5cm,
    minimum height=1.4cm
  }
]

\node[box] (A) at (0,0)
  {Type IIB string theory \\ on resolution
   $\mathbb{R}^{1,5}\times\mathrm{GH}_L $};

\node[box] (B) at (8,0)
  {Type IIB string theory \\ on orbifold $\mathbb{R}^{1,5}\times\mathbb{C}^{2}/\mathbb{Z}_{L }$};

\node[box] (C) at (8,-4)
  {6d $\mathcal{N}=(2,0)$ SUGRA EFT for light modes};

\node[box] (D) at (0,-4)
  {Type IIB SUGRA\\ on resolution  $\mathbb{R}^{1,5}\times\mathrm{GH}_L $};

\draw[->] (A) -- node[above]
  {$a\to 0$} (B);

\draw[->] (B) -- node[right]
  {$\alpha'\to 0$ } (C);
\draw[<-] (C) -- node[above]
  {$a\to0$} (D);
  \draw[->] (A) -- node[right]
  {$\alpha'\to0$} (D);
\end{tikzpicture}
\caption{The resolution of the orbifold singularity provides an interpretation for the massless twisted states. We denote the characteristic scale of the resolution by $a$ and sending it to $0$ we recover the orbifold theory. Similarly, $\alpha'$ controls the mass of excited string states, which become infinitely heavy at $\alpha'\to0$, allowing us to integrate them out of the low-energy effective supergravity theory.  }
\label{fig:typeIIB_limits}
\end{figure}

A geometric interpretation of the massless twisted spectrum of string theory on $\mathbb{R}^{1,5}\cross \mathbb{C}^2/\mathbb{Z}_L$ is attainable via resolving of the orbifold singularity, i.e. by glueing in a geometry that approaches $\mathbb{C}^2/\mathbb{Z}_L$ at large distances from the origin but is completely smooth in the interior. Such resolutions are known as gravitational multi-instanton or Gibbons-Hawking ($\text{GH}_L$) backgrounds \cite{Gibbons:1978tef}. These backgrounds feature non-trivial two-cycles. It turns out that wrapping 10d supergravity fields on these cycles and collecting the moduli of the resolution itself, one can precisely match the massless twisted spectrum of string theory \cite{Douglas:1996sw}. Collapsing the resolution while keeping these fields present provides a rationale for the presence of localised degrees of freedom which can only propagate in six dimensions. This interpretation suggests that the diagram in Figure \ref{fig:typeIIB_limits} commutes. We will review and elaborate on this supergravity construction in Section \ref{sec: 2}.

A similar resolution approach was used in \cite{Gukov:1998kk} to investigate the effective theory of twisted modes in $\mathrm{AdS}_5\cross S^5/\mathbb{Z}_L$, by treating the twisted sector fields as specific modes of 10d supergravity on the resolved background which couple to background fields such as $F_5$, as dictated by the 10d supergravity action. Upon dimensional reduction to 6d, one then ends up with appropriate corrections to flat $\mathcal{N}=(2,0)$ supergravity which capture the non-trivial background geometry. Holographic checks against gauge theory data seem to support this construction \cite{Gukov:1998kk,Billo:2022fnb,Skrzypek:2023fkr}. 

In \cite{Skrzypek:2023fkr}, it was then investigated whether $\alpha'$-corrections to the supergravity theory could be explained via a similar 10d construction and subsequent dimensional reduction to 6d. The $\alpha'$-expansion of the string theoretic Virasoro-Shapiro amplitude generates higher-derivative corrections to 10d supergravity at order $(\alpha')^3$, which feature at least eight derivatives and four fields. The primary example of such a correction is derived by considering a four-graviton amplitude and isolating an effective contribution to the action \cite{Gross:1986iv}
\begin{equation}\label{eq: R4}
    \mathcal{S}_{\rm{SUGRA}}+\alpha'^3\zeta(3) \int \dd^{10} x \sqrt{-g}\left(\mathcal{R}^4+ \dots\right)\,.
\end{equation}
This term has to be completed to a supersymmetric invariant. We take particular note of the factor $\zeta(3)$, which arises in the low-energy expansion of the Virasoro-Shapiro amplitude and is therefore universal to corrections at this order. Inspired by the previous line of reasoning, one may now be tempted to evaluate these corrections for the localised modes of the resolved orbifold background to generate appropriate $(\alpha')^3$-corrections to the 6d supergravity theory for the twisted sector. 

In the simplest case of $\mathrm{AdS}_5\cross S^5/\mathbb{Z}_2$, some suggestive localisation results seem to point towards this possibility. Take for example the (un)twisted half-BPS operators 
\begin{equation}\label{eq: ops}
U_k(y)=\frac{1}{\sqrt{2k }}\left(\frac{2}{N}\right)^{\frac{k}{2}}\Big(\tr \phi_0^k+\tr\phi_1^k\Big)\,,\qquad  T_k(y)=\frac{1}{\sqrt{2k }}\left(\frac{2}{N}\right)^{\frac{k}{2}}\Big(\tr \phi_0^k-\tr\phi_1^k\Big)\,,
\end{equation}
built out of adjoint scalars $\phi_{0,1}$ from the two vector multiplets in the dual theory.
As their conformal dimension $\Delta=k$ is protected, the corresponding 2-point correlators are
\begin{align}\label{eq: two-point}
\langle U_k(y_1)\bar{U}_k(y_2)\rangle&=\frac{G_k(\lambda,N)}{\abs{y_1-y_2}^{2k}}\,, \qquad \langle T_k(y_1)\bar{ T}_k(y_2)\rangle=\frac{R_{k}(\lambda, N)}{\abs{y_1-y_2}^{2k}}\,.
\end{align}
At leading order in large $N$, one finds $G_k(\lambda,N)=1$, which is expected since the untwisted operator descends from $\mathcal{N}=4$ SYM and is canonically normalised. For the twisted sector, on the other hand, $R_{k}$ turns out to be a non-trivial function of $\lambda$ and a strong-coupling expansion yields (see \cite{Galvagno:2020cgq,Beccaria:2021hvt,Billo:2021rdb,Billo:2022gmq} and \cite{Billo:2022fnb,Beccaria:2022ypy,Billo:2022lrv,Beccaria:2023kbl})
\begin{align} \label{eq: Z2-case}
R_{k}(\lambda) =\frac{4\pi^2}{\lambda}k(k-1)\left(\frac{\lambda'}{\lambda}\right)^{k-1}\left[1
+\frac{(2k-1)(2k-2)(2k-3)}{2(\lambda')^{3/2}}\zeta(3) +\order{1/(\lambda')^{5/2}}\right]\,,
\end{align}
where one defines a renormalised coupling $\sqrt{\lambda'}\equiv \sqrt{\lambda}-4\log 2$ \cite{Beccaria:2022ypy,Beccaria:2023kbl}. Of course, one could absorb $R_{k}$ in the normalisation of the twisted sector states, but similar localisation results for three-point functions are governed by normalisation-independent structure functions, which also depend on $\lambda$ in a similar fashion.
In \cite{Billo:2022gmq,Billo:2022fnb}, the leading strong coupling behaviour was matched to the predictions from the low-energy effective action \cite{Gukov:1998kk} for the twisted sector. The subleading correction term scales as $\tfrac{\zeta(3)}{(\lambda')^{3/2}}\sim \zeta(3)(\alpha')^3$, and should thus be matched to higher-derivative corrections such as \eqref{eq: R4}. This was investigated in \cite{Skrzypek:2023fkr}.

In this paper, we want to extend the discussion of the $\mathbb{Z}_2$ orbifold theory analysed in \cite{Skrzypek:2023fkr} to the generic $\mathbb{Z}_L$ orbifold case. The resolution procedure outlined above may be applied to generate an effective action for the massless twisted states in $\mathbb{R}^{1,5}\times \mathbb{C}^2/\mathbb{Z}_L$. An embedding into $\mathrm{AdS}_5\cross  S ^5/\mathbb{Z}_L$ is technically challenging, but can be approximated close to the orbifold singularity. We elaborate on these constructions in Section \ref{sec: 2}. Turning towards $\alpha'$-corrections, we could again employ a reduction of the 10d $(\alpha')^3$-terms to 6d. Intriguingly, localisation results for $\mathbb{Z}_L$ orbifolds suggest that this is not the correct way of constructing $\alpha'$-corrections to the twisted sector EFT. The twisted two-point function \eqref{eq: two-point} now depends on  the twist $n$ of the sector in question. At leading order in large $N$, the normalisation constant takes the form 
\begin{equation}\label{eq: Zl-case}
\begin{split}
&R_{n,k}(\lambda)
=
\frac{4\pi^2}{\lambda}\frac{k(k-1)}{\sin^2\left(\frac{\pi n}{L}\right)}
\left(\frac{\lambda'}{\lambda}\right)^{k-1}
\\[4pt]
&\quad\times
\left[
1
-\frac{(2k-1)(2k-2)(2k-3)}{48(\lambda')^{3/2}}\,
\left[
4\zeta(3)
+\psi^{(2)}\!\left(\frac{n}{L}\right)
+\psi^{(2)}\!\left(1-\frac{n}{L}\right)\right]+\order{1/(\lambda')^{5/2}}\right]\,,
\end{split}
\end{equation}
where the renormalised coupling $\lambda'$ is now defined as
\begin{equation}
\sqrt{\lambda'}\equiv \sqrt{\lambda}+2\gamma+\psi\left(\frac{n}{L}\right)+\psi\left(1-\frac{n}{L}\right)\,,
\end{equation}
which is also twist dependent. In the above, the polygamma function $\psi^{(m)}(x)$ denotes the $m^{\text{th}}$ derivative of the $\Gamma$-function $\psi(x)\equiv\tfrac{\dd}{\dd x}\log\Gamma(x)$, and $\gamma$ denotes the Euler-Mascheroni constant. Crucially, \eqref{eq: Zl-case} is not displaying the universal $\zeta(3)$ factor of the 10d $(\alpha')^3$-term \eqref{eq: R4}. Only for the special values of $L=2,3,4,6$, the digamma functions may be expressed in terms of $\zeta(3)$, but this is a numerological artifact and not representative of the generic case.

This departure from the $\zeta(3)$-scaling suggests that it is too naive to rely on the flat space Virasoro-Shapiro amplitude to generate the $\alpha'$-corrections. In fact, the appropriate string amplitudes generating corrections for twisted sector fields should themselves involve twisted sector vertex operators on the worldsheet. It is furthermore very plausible that a string amplitude involving twisted sector external states would similarly receive contributions from twisted sector states running in the virtual channels. Since the spectrum of twisted sectors has fractional mass-levels (see Appendix \ref{App: Spectrum} for details), the pole structure of such a twisted amplitude differs significantly from that of the usual Virasoro-Shapiro amplitude. In Section \ref{sec: 3}, we explicitly construct a sample NS-sector four-point amplitude involving two twisted and two untwisted massless states and evaluate its low-energy expansion. While the leading term gives the expected supergravity contribution, the first correction at order $(\alpha')^3$ indeed features the polygamma factors observed in \eqref{eq: Zl-case}, demonstrating an impressive matching between localisation and string theory. Our construction furthermore suggests that such polygamma factors are a universal feature of twisted sector amplitudes. 

In Section \ref{sec: 4}, we thus draw the conclusion that despite its success at the supergravity level, the resolution procedure outlined above is not immediately applicable at the level of higher-derivative corrections, where it neglects contributions from virtual twisted sector modes. This was not observed in the $\mathbb{Z}_2$ case \cite{Skrzypek:2023fkr}, due to the fact that $\psi^{(2)}(\tfrac{1}{2})=-14 \zeta(3)$ and the lack of knowledge about the explicit form of the supersymmetric completion of the $\mathcal{R}^4$-term \eqref{eq: R4}\,. We thus require a more complete survey of twisted string amplitudes and the resulting $\alpha'$-corrections to the 6d supergravity theory for a detailed comparison to  the strong-coupling expansion of twisted observables in gauge theory. We plan to return to this challenge in future work.

Finally, we would like to comment on an interesting limit of the $\mathbb{Z}_L$ orbifold theories discussed in this paper, which is the limit of 
$L\to \infty$, where the opening angle of the orbifold singularity shrinks to $0$ while the number of twisted sectors grows infinitely, resulting in an infinitely ``long'' quiver. This limit has recently been considered in the localisation literature \cite{Korchemsky:2025eyc}, and we would like to comment to what level our understanding of $\mathbb{Z}_L$ orbifold theories survives there. These comments are the subject of Appendix \ref{app: largeL}.
\section{The 6d effective action from geometry}\label{sec: 2}
We first review the resolution procedure, which provides a geometric interpretation of the massless twisted sector fields expected from the string theory spectrum (see Appendix \ref{App: Spectrum}). Identifying the twisted sector as localised modes of 10d supergravity, allows us to dimensionally reduce the 10d supergravity action to an effective 6d theory for the local degrees of freedom. One simply has to integrate the respective mode profile across the resolution space. This may be done at an abstract level in algebraic geometry language, where $\mathbb{C}^2/\mathbb{Z}_L$ is known as an $A_{L-1}$ singularity, and where the resolution cycle overlaps are encoded by the elements of the Cartan matrix of the associated Lie algebra in the ADE classification \cite{McKay}. We will review this logic in a more pedestrian  differential geometric language to build intuition for the resolutions at hand.
\subsection{Review of the Gibbons-Hawking background}
We begin by reviewing some properties concerning the resolutions of $\mathbb{C}^{2}/\mathbb{Z}_{L}$ orbifolds. Resolution geometries for this class of singularities are given by the Gibbons-Hawking multi-centered metrics \cite{Gibbons:1978tef}, which take the form
\begin{equation}
\label{GH}
\dd s^{2}_{\mathrm{GH}_{L}}=U^{-1}(x)\big(\dd\tau+w_{i}(x)\dd x^{i}\big)^{2}+U(x)\dd x_{i}\dd x^{i}\,,\qquad U(x)=\sum_{i=1}^{L}\frac{1}{\vert x-\vec{x}_{i}\vert}\,.
\end{equation}
The scalar potential $U(x)$ satisfies the 3d Poisson equation with point charges at $\vec{x}_i$ and the vector potential $w_i(x)$ is related to $U(x)$ via
\begin{equation}
\label{Laplace}
\nabla_{i}U(x)=\pm\varepsilon_{i}^{jk}\nabla_{j}w_{k}(x)\,.
\end{equation}
These metrics are self-dual solutions to the vacuum Einstein equations, and may be described as a non-trivial $\mathrm{U(1)}$ fibration over an $\mathbb{R}^{3}$ base, where the fiber size is controlled by $U(x)^{-1}$ and degenerates at the positions $\vec{x}_{i}$ of the centers (referred to sometimes as gravitational instantons). The vector potential $w_{i}(x)$ defines a $\mathrm{U(1)}$ connection on $\mathbb{R}^{3}$ and has Dirac string-like singularities along lines coming out of each center, which may be removed by appropriate patching around $\vec{x}_{i}$ if the fiber coordinate $\tau$ is taken to have global periodicity $4\pi$ \cite{Gibbons:1978tef}.

Let us first consider a configuration of coincident instantons at $\vec{x}_{i}=0$. Transforming to spherical coordinates, \eqref{GH} becomes \footnote{We have re-scaled the fiber coordinate such that $\psi=\frac{\tau}{L}\in[0,\frac{4\pi}{L})$ to factor out an overall $L$.}
\begin{equation}
\dd s^{2}=\frac{L}{r}\dd r^{2}+4L r\,(\sigma_{x}^{2}+\sigma_{y}^{2}+\sigma_{z}^{2})\,,
\end{equation}
where we have identified the $\mathrm{SU}(2)$ left-invariant Cartan one-forms
\begin{equation}
\label{eq:SU2Cartan1forms}
\sigma_{x}=\frac{1}{2}\big(\sin\psi \,\dd\theta-\sin\theta\cos\psi \,\dd\phi\big)\,,\, \sigma_{y}=-\frac{1}{2}\big(\cos\psi\, \dd\theta + \sin\theta\sin\psi \,\dd\phi\big)\,,\,\sigma_{z}=\frac{1}{2}\big(\dd\psi+\cos\theta \,\dd\phi\big)\,.
\end{equation}
These involve two additional angular coordinates,\footnote{These are $\theta\in[0,\pi]$, $\phi\in[0,2\pi)$. The Cartan one-forms satisfy the useful identity $\dd \sigma_i=\epsilon_{ijk}\sigma_{j}\wedge\sigma_k$.} and allow us to express the three-sphere metric as a Hopf fibration over $S^{2}$. After a coordinate transformation $\rho^{2}=4L r$, we can recast the above metric in the locally flat form
\begin{equation}\label{eq:locallyflatmetric}
\dd s^{2}=\dd\rho^{2}+\rho^{2}(\sigma_{x}^{2}+\sigma_{y}^{2}+\sigma_{z}^{2})\,.
\end{equation}
The case of $L=1$ corresponds to a globally flat metric on $\mathbb{C}^{2}$ in spherical coordinates, while for $L>1$ the metric is globally $\mathbb{C}^{2}/\mathbb{Z}_{L}$ (it is asymptotically locally Euclidean or ALE) . This is due to the (reduced) periodicity $\frac{4\pi}{L}$ of $\psi$, resulting in an orbifold singularity at $\rho=0$. In the limit of coincident gravitational instantons, we thus make contact with the orbifold background.

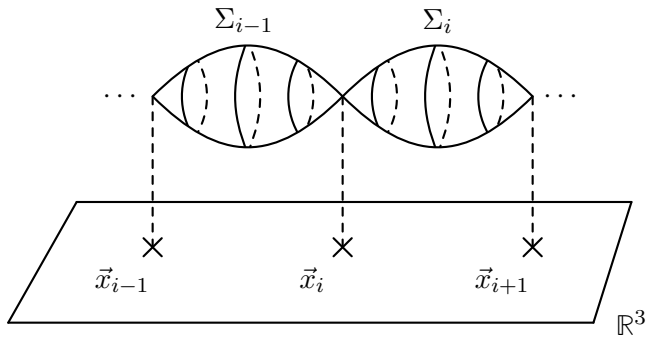
\begin{figure}[ht]
\centering
\begin{tikzpicture}[x=1cm,y=1cm,>=Stealth,line cap=round,line join=round]

  \draw[thick,-] (1,0) -- (8.3,0);
  \coordinate (Xim1) at (2,-0.6);
  \coordinate (Xi)   at (4.5,-0.6);
  \coordinate (Xip1) at (7,-0.6);

\foreach \P/\vlabel/\clabel in {
    Xim1/{$\vec{x}_{i-1}$}/{},
    Xi/{$\vec{x}_{i}$}/{},
    Xip1/{$\vec{x}_{i+1}$}/{}
  }{
    \draw[thick] (\P) ++(-0.12,-0.12) -- ++(0.24,0.24);
    \draw[thick] (\P) ++(-0.12, 0.12) -- ++(0.24,-0.24);

    \node[inner sep=1.5pt,anchor=north]
      at ([xshift=-0.4cm,yshift=-0.20cm]\P) {\vlabel};
    \node[inner sep=1.5pt,anchor=north]
      at ([xshift= 0.34cm,yshift=-0.6cm]\P) {\clabel};
  }

\coordinate (A) at (2.0,1.4);
  \coordinate (B) at (4.5,1.4);
  \coordinate (C) at (7.0,1.4);

  \draw[thick,dashed] (A) -- (Xim1);
  \draw[thick,dashed] (B) -- (Xi);
  \draw[thick,dashed] (C) -- (Xip1);

  \draw[thick,-] (1.0,0) -- (0.1,-1.6);
  \draw[thick,-] (0.1,-1.6) -- (7.8,-1.6);
  \draw[thick,-] (7.8,-1.6) -- (8.3,0);
  \node at (8.3,-1.6) {$\mathbb{R}^{3}$};
  
  \node at (1.6,1.4) {$\cdots$};
  \node at (7.4,1.4) {$\cdots$};

  \coordinate (A) at (2.0,1.4);
  \coordinate (B) at (4.5,1.4);
  \coordinate (C) at (7.0,1.4);

  \def\Leftcycle{
    (A) .. controls (2.9,2.3) and (3.6,2.3) .. (B)
        .. controls (3.6,0.5) and (2.9,0.5) .. (A) -- cycle
  }
  \def\Rightcycle{
  (B) .. controls (5.4,2.3) and (6.1,2.3) .. (C)
      .. controls (6.1,0.5) and (5.4,0.5) .. (B) -- cycle
}

  \draw[thick] \Leftcycle;
  \draw[thick] \Rightcycle;
  \node at (3.2,2.4) {$\Sigma_{i-1}$};
  \node at (5.75,2.4) {$\Sigma_i$};

\def\dx{0.22}   
\def\dy{0.22} 
\begin{scope}
  \clip \Leftcycle; 

  \foreach \x/\ylo/\yhi [
    evaluate=\ylo as \ymid using {0.5*(\ylo+\yhi)}
  ] in {
    2.55/0.85/1.95,
    3.25/0.65/2.15,
    3.95/0.85/1.95
  }{
    \draw[thick]
      (\x,\ylo) .. controls (\x-\dx,\ymid-\dy) and (\x-\dx,\ymid+\dy) .. (\x,\yhi);
    \draw[thick,dashed]
      (\x,\ylo) .. controls (\x+\dx,\ymid-\dy) and (\x+\dx,\ymid+\dy) .. (\x,\yhi);
 }
\end{scope}

\begin{scope}
  \clip \Rightcycle;

  \foreach \x/\ylo/\yhi [
    evaluate=\ylo as \ymid using {0.5*(\ylo+\yhi)}
  ] in {
    5.05/0.85/1.95,
    5.75/0.65/2.15,
    6.45/0.85/1.95
  }{
    \draw[thick]
      (\x,\ylo) .. controls (\x-\dx,\ymid-\dy) and (\x-\dx,\ymid+\dy) .. (\x,\yhi);

    \draw[thick,dashed]
      (\x,\ylo) .. controls (\x+\dx,\ymid-\dy) and (\x+\dx,\ymid+\dy) .. (\x,\yhi);
  }
\end{scope}
\end{tikzpicture}
\caption{Depiction of the resolution space \eqref{GH} as an $S^{1}$ fibration over $\mathbb{R}^{3}$ for a generic configuration of instantons at $\vec{x}_{i},\vec{x}_{i\pm 1}\in\mathbb{R}^{3}$. The two-cycle $\Sigma_{i}$ is spanned by the fiber and a curve connecting $\vec{x}_{i},\,\vec{x}_{i+1}$.}
\label{fig:overlap-cycles}
\end{figure}

The orbifold singularity is resolved when the gravitational instantons become spatially separated ($\vec{x}_i\neq\vec{x}_j$ if $i\neq j$), allowing for the construction of a smooth atlas. The relative positions of the $L$ centers on the base then generate $3(L-1)$ genuine moduli of the resolution space. Because the fiber shrinks to a point at the positions of the instantons, one expects finite-sized two-cycles $\{\Sigma_{n}\}$ arising as $S^{1}$ fibrations over line segments connecting any two centers. One may convince oneself that the number of linearly independent two-cycles is $(L-1)$. This configuration is depicted in Figure \ref{fig:overlap-cycles}.

When studying supergravity on such resolved backgrounds, we may thus consider form-fields wrapping these two-cycles. In particular, there exists a set of $(L-1)$ anti-self-dual, closed, normalisable two-forms $\{\omega_{n}\}$ which are representatives of the non-trivial cohomology classes of $H^{2}(\mathrm{GH}_{L},\mathbb{Z})$. We choose to normalise them as
\begin{equation}
\label{orthogonal}
\oint_{\Sigma_{n}}\omega_{m}=\delta_{nm}\,.
\end{equation}
The normalisability property of the two-forms $\omega_n$ implies that their overlap over the resolution space is finite. A direct calculation of the overlap integral (carried out in the following section \eqref{sect:effectiveaction}) reveals that the overlap integral is given by
\begin{equation}
\label{overlapint}
\int_{\mathrm{GH_{L}}}\omega_{n}\wedge\star\omega_{m}=-(C^{-1})_{nm}\,,
\end{equation}
where $(C^{-1})_{nm}$ denote the elements of the inverse Cartan matrix of the $A_{L-1}$ algebra. Poincar\'e duality then relates \eqref{overlapint} to the intersection matrix of a set of two-cycles $\{\Sigma_{n}\}$, which form a basis of representatives of the non-trivial homology classes in the Poincar\'e dual $H_2(\mathrm{GH}_{L},\mathbb{Z})$, as $\Sigma_n\cap\Sigma_m=-(C^{-1})_{nm}$.\footnote{The canonical intersection matrix of the two-cycles on the $\mathrm{GH}_{L}$ resolution is, however, given by $-C_{nm}$ \cite{McKay}. The two-forms $\{\tilde{\omega}_{n}\}$ which are Poincar\'e dual to the set of canonically-intersecting two-cycles are related to the two-forms in \eqref{orthogonal} via the change of basis $\tilde{\omega}_{n}=-C_{mn}\,\omega_{m}$.} 
The basis of two-forms defined by \eqref{orthogonal} and \eqref{overlapint} was the one used in \cite{Klebanov:1999rd,Billo:2022fnb} to construct an effective action for twisted scalar modes on $\mathrm{AdS}_5\cross S ^1$, which was then used to match localisation results of the dual quiver gauge theory. We will henceforth adopt this basis.
\subsection{Effective action for wrapped two-form fields}\label{sect:effectiveaction}
With Figure \ref{fig:typeIIB_limits} in mind, let us now consider type IIB supergravity on the resolved geometry. The ten-dimensional supergravity fields wrapping the two-cycles, in combination with the resolution moduli, generate the expected $(L-1)$ tensor multiplets of 6d $\mathcal{N}=(2,0)$ supergravity. The bosonic field content can be understood as follows: the wrapped RR four-form $C_4$ becomes a self-dual two-form in the 6d effective theory, while the wrapped NSNS and RR two-forms $B_2$ and $C_2$ become 6d scalars $b_n(y)$ and $c_n(y)$ via the decomposition \footnote{Unless otherwise specified, we will collectively denote by $y$ the coordinates of the `external' factor of the target space (here $\mathbb{R}^{1,5}$), and by $x$ the coordinates of the `internal' factor (here $\mathbb{C}^{2}/\mathbb{Z}_L$).}
\begin{equation}\label{eq:B2decomp}
    B_2=\sum_{n=1}^{L-1}b_n(y)\omega_n(x)\,,\qquad C_2=\sum_{n=1}^{L-1}c_n(y)\omega_n(x)\,. 
\end{equation}
Each resolution cycle further contributes three geometric moduli, completing the tally of bosonic modes in a tensor multiplet of 6d $\mathcal{N}=(2,0)$ supergravity. Focusing on the wrapped $B_2$-field, which, for $L=2$, corresponds to operators $T_k(y)$ in \eqref{eq: ops}, the idea is to obtain explicit field configurations and formulate their low-energy effective action.

We begin by arguing that the effective action for wrapped $B_2$-field is insensitive to the configuration of instantons on the base. To see this, we assume an arbitrary configuration of non-coincident instantons on $\mathbb{R}^{3}$. Two-cycles can be created by lifting curves $\gamma: [0,1]\to\mathbb{R}^{3}$ with endpoints $\gamma(0)=\vec{x}_{i}$ and $\gamma(1)=\vec{x}_{j}$ to the total space (including the fiber coordinate). If the endpoints coincide $(\vec{x}_{i}=\vec{x}_{j})$, then $\gamma(t)$ is a closed curve on $\mathbb{R}^{3}$ and can be contracted to a point. In this case, the two-cycle is homologically trivial and can be regarded as the boundary of a surface in $\mathbb{R}^{3}$. To construct $H_{2}(\mathrm{GH}_{L},\mathbb{Z})$, we must identify non-trivial two-cycles up to such boundary cycles. This means that all curves $\gamma(t)$ with equal endpoints are homologically equivalent. Using this equivalence relation, we conclude that any set of linear-independent non-trivial elements of $H_{2}(\mathrm{GH}_{L},\mathbb{Z})$ forms a maximal tree of $(L-1)$ edges connecting the different centers $\vec{x}_{i}$. Different maximal trees thus correspond to equivalent bases of representatives (up to an isomorphism) of the non-trivial homology classes of $H_{2}(\mathrm{GH}_{L},\mathbb{Z})$.

$B_2$ is defined up to gauge transformations $B_{2}\to B_{2}+\dd\chi_{1}$, with $\chi_{1}$ a globally well-defined one-form. This gauge freedom allows us to set $B_{2}=0$ on trivial cycles. However, $B_{2}$ may still have (non-exact) components in $H^{2}(\mathrm{GH}_{L},\mathbb{Z})$, from which it follows that its periods $b_{n}=\oint_{\Sigma_{n}}B_{2}$, must be gauge invariant.\footnote{This can be seen explicitly by combining a gauge transformation with Stoke's theorem so that $b_{n}\to\, b_{n}+\oint_{\Sigma_{n}}\dd\chi_{1} =\,b_{n}+\oint_{\partial\Sigma_{n}}\chi_{1}=\,b_{n}$.} Furthermore, any deformation of the cycle $\Sigma_n$ can be written as the addition of an exact two-cycle to $\Sigma_n$, on which $B_2$ can be gauged away. The $B_{2}$-field wrapping these cycles is therefore not sensitive to the actual embeddings, only to the topology of the maximal tree, thus generating the expected $(L-1)$ twisted scalars $b_{n}$ in the resolved geometry. We adopt a naming convention for the wrapped $B_2$-field components where we choose an ordering of instantons such that the periods $b_{n}$ are given by the component of the total $B_{2}$-field wrapping the two-cycle with endpoints at $\vec{x}_{n}$ and $\vec{x}_{n+1}$.

Upon integrating the 10d supergravity action on such wrapped cycles, we end up with overlap integrals of the form \eqref{overlapint}, which are topological invariants. When considering $b_n(x)$ and $c_n(x)$ as in \eqref{eq:B2decomp}, we may therefore distribute the instanton centers $\vec{x}_i$ in whatever configuration we like without changing the outcome.\footnote{This may also be checked explicitly, e.g. by varying the position of an instanton and confirming invariance of the effective action.} For concreteness, we will therefore choose to work with a configuration of non-coincident, equally-spaced instantons along the $z$-axis in cylindrical coordinates $(r,\varphi,z)$ on the base. In these coordinates, the relevant quantities of the metric \eqref{GH} are
\begin{equation}\label{eq:cylindricalcoords}
    U(r,z)=\sum_{i=1}^{L}\frac{1}{\sqrt{r^2+(z-z_i)^2}}\,,\qquad w_i(r,z)\dd x^i=\sum_{i=1}^{L}\frac{(z-z_i)}{\sqrt{r^2+(z-z_i)^2}}\dd\varphi\,.
\end{equation}
A basis of non-trivial 2-cycles $\Sigma_n$ is spanned by the fiber $\tau$ and the line segments $z\in[z_i,z_{i+1}]$ at $r=0$. 

We now present the explicit $B_{2}$-field configuration wrapping these resolution cycles $\Sigma_n$. The appropriate closed, anti-self-dual two-forms may be expressed as \footnote{In cylindrical coordinates, we take the orthonormal frame of the metric \eqref{GH} to be $e^{1}=\sqrt{U}\dd r\,,\; e^{2}=r\sqrt{U}\dd\varphi\,,\; e^{3}=\sqrt{U}\dd z\,,\; e^{4}=\tfrac{1}{\sqrt{U}}(\dd\tau+w_{\varphi}\dd\varphi)$, with orientation $\varepsilon_{1234}=+1$.}
\begin{equation}
\label{eq:GH2normalisedB2field}
\omega_n(r,z)=\mathcal{A}_n(r,z)(e^{1}\wedge e^{2}-e^{3}\wedge e^{4})+\mathcal{B}_n(r,z)(e^{2}\wedge e^{3}-e^{1}\wedge e^{4})\,,
\end{equation}
where the functions $\mathcal{A}_n(r,z)$ and $\mathcal{B}_n(r,z)$ are given by
\begin{align} \label{eq:B2components}&\mathcal{A}_n(r,z)=\frac{W_n^2-U^2}{16\pi}\left[\frac{zW_n}{U}+\frac{1}{2}\bigg(1+\frac{W_n^2}{U^2}\bigg)\right]\,,\\
&\mathcal{B}_n(r,z)=\frac{W_n^2-U^2}{16\pi}\frac{2rW_n}{U}\,.
\end{align}
In the above, we have introduced the ``auxiliary'' scalar potential
\begin{equation}
\label{eq:arbLauxpot}
W_{n}(r,z)=\sum_{i=1}^{L}\frac{c_{i,n}}{\sqrt{r^2+(z-z_i)^2}}\,,\quad c_{i,n}=\begin{cases}
-1,\; i\leq n\\
1,\;\;\;\, i> n
\end{cases}\,,
\end{equation}
which is an ordered sum with coefficients $\pm 1$. Since $\omega_n$ is closed, it is locally exact. Away from $r=0$, it may therefore be expressed through the one-form potential
\begin{equation}
\chi_{n}(r,z)=\frac{1}{8\pi}\Bigg[\frac{1}{r}\frac{\mathcal{W}_{n}(r,z)}{\sqrt{U(r,z)}}e^2-\frac{W_{n}(r,z)}{\sqrt{U(r,z)}}e^4\Bigg]\,,\qquad \mathcal{W}_{n}(r,z)=\sum_{i=1}^{L}c_{i,n}\frac{(z-z_{i})}{R_{i}}\,,
\end{equation}
such that $\omega_n=\dd\chi_n$. Here, we defined the ``auxiliary'' connection $\mathcal{W}_{n}(r,z)$, which obeys the relation \eqref{Laplace} with the auxiliary scalar potential \eqref{eq:arbLauxpot}.

We now need to determine various integrals of the constructed two-forms $\omega_n$. First, the period integral for any component $\omega_{n}$ wrapping the $i^{th}$ two-cycle $\Sigma_{i}$ is given by
\begin{equation}\label{eq:periodintegrals}
\oint_{\Sigma_{i}}\omega_{n}=\,\frac{c_{i+1,n}-c_{i,n}}{2}=\delta_{i,n}\,,
\end{equation}
obeying precisely the required normalisation \eqref{orthogonal}. One may then check explicitly that the overlap integrals
\begin{equation}
\label{eq:GHL2formOverlaps}
\int_{\mathrm{GH_{L}}}\omega_{n}\wedge \star\omega_{m}=-(C^{-1})_{mn}=\frac{mn}{L}-\mathrm{min}\{m,n\}\,,
\end{equation}
are satisfied, as expected from algebraic geometry (cf. \eqref{overlapint}). 

We are now in a position to derive the effective 6d dynamics of the type IIB twisted scalars $b_n(y)$, corresponding to the wrapped components of $\omega_{n}$ on the resolution cycles of $\mathbb{R}^{1,5}\times\mathbb{C}^{2}/\mathbb{Z}_{L}$. The relevant part of the ten-dimensional type IIB supergravity action is
\begin{equation}
\label{eq:10daction}
S_{10}\supset-\frac{1}{2\kappa_{10}^{2}}\int \dd^{10}x \sqrt{g_{10}}\bigg(\frac{1}{2\cdot3!}\vert H_{3}\vert ^{2}\bigg)\,,\qquad H_{3}=\dd B_{2}\,,
\end{equation}
where $2\kappa^{2}_{10}=(2\pi)^{7}g_{s}^{2}(\alpha')^{4}$ is the ten-dimensional gravitational constant. The integration over $\mathrm{GH}_{L}$ boils down to \eqref{eq:GHL2formOverlaps}, but we will restore appropriate length units and decompose the total $B_{2}$ field as $B_{2}(y,x)=(4\pi^{2}\alpha')\sum_{n}b_{n}(y)\omega_{n}(x)$, where $x$ now collectively denotes the coordinates of $\mathrm{GH}_{L}$. Performing the integral, we obtain the following effective six-dimensional action for the twisted scalars $b_{n}(y)$
\begin{equation}
\label{eq:discrete6daction}
\begin{split}
S_{\rm{eff}}&\supset\frac{1}{2\kappa_{6}^{2}}\int_{\mathbb{R}^{1,5}} \dd^{6}y\sqrt{-g_{6}}\,\frac{1}{2}\sum_{n,m=1}^{L-1}(C^{-1})_{nm}\nabla b _{n}(y)\cdot\nabla b_{m}(y) \,,\\
\frac{1}{2\kappa_{6}^{2}}&=\frac{(4\pi^{2}\alpha')^{2}}{2\kappa_{10}^{2}}=\frac{1}{(2\pi)^{3}g_{s}^{2}(\alpha')^{2}}\,,
\end{split}
\end{equation}
where the mixing between different scalar modes is a consequence of the non-trivial overlaps \eqref{eq:GHL2formOverlaps}. The same effective action would also arise for the twisted scalars $c_{n}(y)$, descending from the RR two-form $C_2$.
\subsection{Towards embedding in $\mathrm{AdS}_5\cross  S ^5$}

So far, we have considered supergravity on the Ricci-flat resolution space $\mathbb{R}^{1,5}\times \mathrm{GH}_{L}$ and constructed a 6d effective theory for the twisted sector fields propagating in $\mathbb{R}^{1,5}$. In the context of holography, one would instead like to study the resolution of $\mathrm{AdS}_5\cross S^5/\mathbb{Z}_{L}$, which is supported by $F_5$-flux. 

In \cite{Gukov:1998kk}, an approximation close to the orbifold singularity was considered, in which the space-time approaches $\mathrm{AdS}_5\cross S^1\cross \mathbb{C}^2/\mathbb{Z}_{L}$. The argument was simply to consider the 10d supergravity action in the background $\mathrm{AdS}_5\cross S^1\cross \mathrm{GH}_{L}$ and analyse what kind of coupling to the background could arise for the wrapped supergravity modes which constitute the twisted sector. For the wrapped $B_2$ and $C_2$ fields discussed above, a coupling to the background $F_5$-flux is mediated by the Chern-Simons term in the 10d action 
\begin{equation}
S_{10}\supset-\frac{1}{4\kappa_{10}^2} \int \, B_2\wedge F_3\wedge F_5\,,\qquad F_3=\dd C_2\,,
\end{equation}
which, evaluated on the approximate geometry and after collapsing the resolution, yields a contribution 
\begin{equation}\label{eq: backgroundcoupling}
S_{\rm{eff}}\supset\frac{1}{2\kappa_{6}^{2}}\int_{\mathrm{AdS}_5\times S^1} \dd^{6}y\sqrt{-g_{6}}\left(\,-4\sum_{n,m=1}^{L-1} (C^{-1})_{nm}b _{n}(y)\cdot \partial_\chi c_{m}(y)\right) \,.
\end{equation}
Here, $\partial_\chi$ is the derivative in the $S^1$-direction. This contribution should be added to the quadratic action \eqref{eq:discrete6daction} (and the equivalent one for $c_n(y)$), which, in turn, should be evaluated on $\mathrm{AdS}_5\cross S^1$. The $b_n(y)$ and $c_n(y)$ fields can be further decomposed into Fourier modes on $S^1$, labelled by integers $k$. Finally, diagonalisation of the kinetic term by introducing $a^{\pm}_n=c_n\pm i b_n$ \cite{Gukov:1998kk,Billo:2022fnb, Skrzypek:2023fkr} results in the $\mathrm{AdS}_5$ spectrum
\begin{equation}
    m_\pm^2=k(k\pm 4)\,,
\end{equation}
which matches the spectrum observed in the dual gauge theory. This justifies using 10d supergravity in the approximate $\mathrm{AdS}_5\cross S^1\cross \mathrm{GH}_{L}$ background to generate a 6d (5d after KK-reduction on $S^1$) effective theory at least at two-derivative level.  

In order to compare to more refined quantities such as correlators like \eqref{eq: two-point}, which localisation can compute in principle to any order in large $\lambda$, one would need to extend the analysis beyond supergravity to string theory corrections. In 10d, these take the form of higher-derivative corrections starting at order $(\alpha')^3$ (cf.\eqref{eq: R4}). Similar corrections are expected in lower dimensions. Indeed, in the strong coupling expansion in \eqref{eq: Zl-case}, the first correction term (after renormalisation of $\lambda$, which was discussed in \cite{Beccaria:2022ypy}) appears at order $(\lambda')^{-\frac{3}{2}}\sim(\alpha')^3$. This is in contrast to the unorbifolded $\mathrm{AdS}_5\cross S^5$ background, where such corrections are generically absent, matching a direct evaluation of the $\mathcal{R}^4$-term \eqref{eq: R4} which vanishes as well. Thus, the orbifold singularity itself seems to cause $(\alpha')^3$-corrections to appear.

In an ideal world, one would like to perform a string theory calculation on the $\mathrm{AdS}_5\cross S^5/\mathbb{Z}_{L}$ background to directly match the $(\alpha')^3$-corrections predicted by localisation. Quantisation of string theory on RR-backgrounds such as $\mathrm{AdS}_5\cross S^5$ has not been successfully implemented yet, so a quantisation of string theory in the orbifold background is clearly beyond reach. We can, however, quantise string theory in the flat background $\mathbb{R}^{1,5}\cross \mathbb{C}^2/\mathbb{Z}_{L}$ and try to at least attain some structural matching. This will be the approach presented in Section \ref{sec: 3}. 

An alternative approach would be to rely on the reasoning in Figure \ref{fig:typeIIB_limits}, also for $\mathrm{AdS}_5\cross S^5/\mathbb{Z}_{L}$. This would entail finding a smooth string background $\mathrm{AdS}_5\cross \mathcal{M}^5_{L}$ that resolves the orbifold singularity and allows us to study 10d supergravity and its $\alpha'$-corrections, in particular for twisted sector fields wrapping (or parameterising) the resolution cycles. This would be a direct refinement of the approximation in \cite{Gukov:1998kk}, which only considers the local $\mathrm{AdS}_5\cross S^1\cross \mathrm{GH}_{L}$ geometry. For $\mathbb{Z}_2$ orbifolds, such a resolution was constructed in \cite{Skrzypek:2023fkr}, albeit featuring some irregular monodromies. This could be achieved since $\mathrm{GH}_{2}$ is equivalent \cite{Prasad:1979kg} to the Eguchi-Hanson space \cite{Eguchi:1978gw,Eguchi:1978xp}, which allows for a simple introduction of background curvature. For $L\neq 2$, no such coordinate system is available and resolutions $\mathcal{M}^5_{L}$ are unknown. 

As a first step towards constructing these resolutions, one can instead consider the limit $k\to\infty$ with fixed $\nu\equiv\frac{k}{\sqrt{\lambda}}$, in which the strings under consideration move with large angular momentum on the $S^1$ subspace of the orbifold fixed locus. In the moving frame of such strings,  the ambient $\mathrm{AdS}_5\cross S^5/\mathbb{Z}_{L}$ approaches a pp-wave background \cite{Blau:2001ne, Berenstein:2002jq,Itzhaki:2002kh, Alishahiha:2002ev, Kim:2002fp, Floratos:2002uh, Sahraoui:2002sp} described by the metric
\begin{align}
\dd s_{10}^2=&- 4\dd y^+\dd y^- - \mu^2\big(y^2+\rho^2\big)(\dd y^+)^2+ \dd y^i\dd y^i + \,\dd \rho^2+\rho^2\big(\sigma_x^2 + \sigma_y^2+ \, \sigma_z^2\big) \,,
\end{align}
where $y^\pm$ are light-cone coordinates built from $\mathrm{AdS}$-time and the $S^1$-angle $\chi$. The coordinates $y^i$ ($i=1,2,3,4$) originate from the remaining $\mathrm{AdS}_5$ dimensions and we used the parameterisation of \eqref{eq:locallyflatmetric} to parameterise $\mathbb{C}^2/\mathbb{Z}_L$. This background is a pp-wave with the transverse geometry $\mathbb{R}^4\times \mathbb{C}^2/\mathbb{Z}_L$. The supporting $5$-form flux becomes 
\begin{equation}\label{eq: flux}
F_{5}= 4\mu\, \dd y^+\wedge\big( \dd y^1\wedge \dd y^2\wedge \dd y^3\wedge \dd y^4 - \rho^3 \dd \rho \wedge \sigma_x\wedge \sigma_y\wedge \sigma_z\big)\,.
\end{equation}
This geometry can be explicitly resolved by glueing in a Gibbons-Hawking geometry \eqref{GH}
\begin{align}\label{eq: ppwavesol}
\dd s_{10}^2=&- 4\dd y^+\dd y^- - \mu^2\big(y^2+h(x,\tau)\big)(\dd y^+)^2+ \dd y^i\dd y^i + \dd s^{2}_{\mathrm{GH}_{L}} \,,
\end{align}
where $h(x,\tau)$ is a solution of the Poisson equation
\begin{equation}
\nabla^2_{_{\mathrm{GH}_{L}}} h(x,\tau)= 8\, . 
\end{equation}
The simplest particular solution is given by
\begin{equation}
\label{eq:ppwavefiniteL}
h(x,\tau)=4\sum_{n=1}^{L}\vert x-\vec{x}_{i}\vert\,.
\end{equation}
We could add harmonic pieces $h_{0}(x,\tau)$ satisfying $\nabla^2_{_{\mathrm{GH}_{L}}} h_0=0$ to this solution, but will refrain for the time being.\footnote{The appropriate choice of harmonic piece would presumably be fixed by taking the Penrose limit of the full resolution $\mathrm{AdS}_5\times \mathcal{M}^5_{L}$.} For any such choice, the resolved pp-wave background \eqref{eq: ppwavesol} constitutes a solution of the supergravity equations of motion once the flux is updated to 
\begin{equation}
F_{5}= 4\mu\, \dd y^+\wedge\big( \dd y^1\wedge \dd y^2\wedge \dd y^3\wedge \dd y^4 - \text{vol}_{\mathrm{GH}_{L}})\,.
\end{equation}
We would now like to investigate the wrapped $B_2$ and $C_2$ modes with large angular momentum $k$ along the $S^1$ subspace. The naive Ansatz \eqref{eq:B2decomp} has to be modified to
\begin{equation}\label{eq: wrapped}
B_{2}(y^\pm,y, r,z)=e^{2i\nu y^-}\sum_{n=1}^{L-1}b_{n}(y)(\omega_n+\mu \Delta\omega_{n,\nu})\,,
\end{equation}
in order to solve the equations of motion. The correction term $\Delta \omega_{n,\nu}$ features mixed terms of the form $\dd y^+\wedge e^i$ but does not functionally depend on the flat coordinates $(y^+,y_i)$. The fields $b_n(y)$ and $c_n(y)$ again mix due to the interaction term \eqref{eq: backgroundcoupling}. After combining $a^{\pm}_n=c_n\pm i b_n$ as before, we find the scalar equation of motion 
 \begin{equation}\label{eq: ppEoM}
     \left[2i\nu\partial_++\partial_i\partial^i+2i\nu\partial_+-\mu^2 \nu^2 \abs{y}^2\mp4\mu \nu\right]a^{\pm}_n(y)=0\,,
 \end{equation}
 appropriate for a pp-wave background with an interaction term \eqref{eq: backgroundcoupling}. The main technical challenge is the explicit construction of $\Delta\omega_{n,\nu}$, which has been successful in the $\mathbb{Z}_2$ case (see \cite{Skrzypek:2023fkr}) but involves the solution of intricate systems of partial differential equations, going beyond the scope of this paper. 

 With this setup in place, one may now consider $(\alpha')^3$-corrections to the 10d supergravity action. The complete supersymmetric completion of the $\mathcal{R}^4$ term \eqref{eq: R4} is not known at this time (see  \cite{Green:1982sw,Gross:1986iv,Sakai:1986bi,Grisaru:1986vi,Freeman:1986zh,Liu:2013dna,Liu:2019ses,Liu:2022bfg} for partial results). Instead, \cite{Skrzypek:2023fkr} considered general covariant eight-derivative terms involving four-fields, two of which being $G_3=F_3+i H_3$ and the other two being non-vanishing background fields on \eqref{eq: ppwavesol}. Focusing on the terms with the highest number of derivatives acting on $G_3$, the following index structure may appear in the 10d Lagrangian \footnote{We denote ten-dimensional indices by greek letters, e.g. $\mu\in\{0,1,2,...,9\}$.} 
 \begin{equation}\label{eq: correction}
\mathcal{L}\supset \ {(\alpha')^3} \zeta(3) 
F^{\mu\nu\sigma\tau\rho}\mathcal{C}_{\tau\rho}\,^{\alpha\beta}\ \Big( i\nabla_\mu \bar G_{\nu\sigma}\,^\gamma \nabla^2 G_{\alpha\beta\gamma} + \dots \Big)\,.
\end{equation}
Here, $\mathcal{C}$ denotes the Weyl tensor built from derivatives of the metric. Inserting the wrapped fields as in \eqref{eq: wrapped} and integrating this term over the resolution space $\mathrm{GH}_{L}$, one ends up with a quite complicated correction to the 6d effective theory on $(y^\pm,y_i)$-space. In the limit where the resolution shrinks to a point $\abs{\vec{x_i}}\sim a\to 0$, the Weyl tensor has divergent components, which can be cancelled by suitable linear combinations with subleading terms in \eqref{eq: correction} \cite{Skrzypek:2023fkr}. The resulting finite part then contributes 
\begin{equation}
 \Delta S_6\ \sim \ \zeta(3) \, 
\int\dd y^+\dd y^-\dd^4 y\,
\,\sqrt{-g_6}\,\sum_{n,m=1}^{L-1} (C^{-1})_{nm} \Big( i \nabla^\mu\bar{a}_n \nabla_-^3 \nabla_\mu a_m + 
 4 \bar{a}_n \nabla_-^4 a_m+ \text{c.c.}\Big)\,.
\end{equation}
This term results in a correction of the two-point function of the twisted fields without changing the conformal dimension \footnote{The argument in \cite{Skrzypek:2023fkr} 
leading to this result did not rely on the precise form of $\Delta \omega_{n,k}$.}
\begin{equation}
    R_{n,k}\sim1+ \zeta(3)\nu^3 c\,,
\end{equation}
where a numerical constant $c$ was left undetermined.

In order to compare this correction to that predicted by the localisation result \eqref{eq: Zl-case}, we take a BMN-type limit of \eqref{eq: Zl-case} where $k\to\infty$ with fixed $\nu\equiv\frac{k}{\sqrt{\lambda}}$ at strong coupling. The precise form of the correction as predicted by localisation is then
\begin{equation}\begin{split}
\label{eq:BMNexpansion2ptfunct}
R_{n,k}(\lambda)\to \frac{4\pi^2\nu^2}{\sin^2\left(\frac{n\pi}{L}\right)}e^{2\nu\left(\sqrt{\lambda'}-\sqrt{\lambda}\right)}\Bigg[1-\frac{\nu^3}{6}\left[
4\zeta(3)
+\psi^{(2)}\!\left(\frac{n}{L}\right)
+\psi^{(2)}\!\left(1-\frac{n}{L}\right)\right]+\order{\nu^5}\Bigg]\,.
\end{split}\end{equation} 

Comparing these expressions we can indeed confirm that the kinematic factor $\nu^3$ is reproduced. However, we observe a disparity in the transcendental factor. The localisation result features polygamma terms that are not present in the naive dimensional reduction of the 10d $(\alpha')^3$ term, which has a universal scaling of $\zeta(3)$. The polygammas may be expressed in terms of $\zeta(3)$ for $L= 2,3,4,6$, but in the generic case no such simplification exists, resulting in a concrete mismatch.

This calls into question whether the 10d reasoning presented in Figure \ref{fig:typeIIB_limits} applies at the level of $(\alpha')^3$-corrections. Our approach so far was to expand around $\alpha'\to 0$ and then take the resolution size $a$ to $0$, but one may doubt whether this is the correct order to consider. 

\section{The twisted Virasoro-Shapiro amplitude}\label{sec: 3}
We now consider the opposite order of limits in Figure \ref{fig:typeIIB_limits}, where we first collapse the resolution to the honest orbifold background $\mathbb{R}^{1,5}\times\mathbb{C}^{2}/\mathbb{Z}_{L }$ and then study type IIB theory in this background. The first step is a computation of the spectrum, which we review in Appendix \ref{App: Spectrum}. We then want to compute string amplitudes, but in order to account for untwisted and twisted external states, we need to modify the vertex operators at play. In particular, the twisted sector vertex operators cannot feature macroscopic momentum in the $\mathbb{C}^{2}/\mathbb{Z}_{L}$ plane, and are therefore localised to the orbifold singularity. 

Once the string amplitudes have been calculated, one may perform an expansion in small $\alpha'$ and consider the effective theory governing the massless spectrum of the superstring. In regular type IIB theory, the large amount of supersymmetry constrains this effective field theory severely, leading to 10d type IIB supergravity. This supergravity theory captures exactly the tree-level two- and three-point correlators of string theory with massless external particles. However, already at the level of four-point functions, additional $(\alpha')^3$-corrections emerge, which result in eight-derivative correction terms such as \eqref{eq: R4} to this supergravity theory \cite{Gross:1986iv}. In the case of the orbifold theory, only half-maximal supersymmetry is preserved. Moreover, since the twisted sector is localised on a 6d submanifold of the target space, the natural language is that of 6d $\mathcal{N}=(2,0)$ supergravity with one supergravity and $(L+1)$ tensor multiplets. This theory will also receive $\alpha'$-corrections from four-point string amplitudes.  

In order to determine the appropriate $\alpha'$-corrections from string theory, we have a plethora of twisted correlators to compute, additional to the untwisted ones we can borrow from the regular type IIB theory. Twisted correlators have been discussed in \cite{Hamidi:1986vh, Dixon:1986qv}. Here, we will present an example of such a calculation and show that the structure of $(\alpha')^3$-corrections in orbifold theories differs from the conventional type IIB expectations. In particular, we observe a departure from the universal $\zeta(3)$ prefactor in type IIB theory towards a combination of polygamma values. We expect this non-trivial transcendental structure to be a universal feature of twisted amplitudes. This matches the observation made in the localisation literature (cf. \eqref{eq: Zl-case}), demonstrating impressive consistency across the holographic duality. 

\subsection{Construction of a sample amplitude}

In order to compute the necessary string amplitude, we choose to work in the RNS formalism. In flat space, the worldsheet theory is governed by the action 
\begin{equation}
    \mathcal{S}_{\text{RNS}}=-\frac{1}{2\pi}\int \dd z\dd \bar{z}\left(\frac{2}{\alpha'}\partial X^\mu \bar\partial X_\mu+i\psi^\mu \bar\partial\psi_\mu +i \tilde\psi^\mu \partial \tilde\psi_\mu  \right)\,,
\end{equation}
where we chose super-conformal gauge. The relevant OPEs are 
\begin{equation}
    \partial X^\mu(z_1) \partial X^\nu(z_2)\sim-\frac{\alpha'}{2}\frac{\eta^{\mu\nu}}{(z_1-z_2)^2}\,,\qquad \psi(z_1) ^{\mu}\psi(z_2)^{\nu} \sim \frac{\eta^{\mu\nu}}{z_1-z_2}\,,
\end{equation}
and similarly for the anti-holomorphic fields. The bosons can be coherently superimposed to form vertex operators with OPE
\begin{equation}\label{eq: VOPE}
    :e^{i k_1^\mu X_\mu(z_1,\bar{z}_1)}::e^{i k_2^\mu X_\mu(z_2,\bar{z}_2)}:\,\sim \abs{z_1-z_2}^{\alpha' k_1\cdot k_2}:e^{i (k_1^\mu+k_2^\mu) X_\mu(z_2,\bar{z}_2)}:\,.
\end{equation}

We will also choose to bosonise the fermions, such that they can be represented in terms of five complex scalar fields $H_i(z)$ ($i\in\{1,2,3,4,5\}$) as
\begin{equation}\label{eq:bosonisedfermions}
    \frac{1}{\sqrt{2}}(\psi^{2i-1}(z)\pm i \psi^{2i}(z))\sim\, :e^{\pm i H_i(z)}:\,.
\end{equation}
To ensure consistent OPEs, we normalise $H_i$ such that 
\begin{equation}
    \partial H_i(z_1) \partial H_j(z_2)\sim-\frac{\delta_{ij}}{(z_1-z_2)^2}\,,\qquad :e^{ i H_i(z_1)}::e^{- i H_i(z_2)}:\sim\frac{1}{z_1-z_2}\,.
\end{equation}
Analogously, the anti-holomorphic fermions are bosonised to yield the anti-holomorphic part of $H_i$.\footnote{We neglect subtleties about bosonic zero-modes and will, in the following, often drop the worldsheet-coordinate dependence for brevity.}

In order to perform the orbifolding procedure, we need to identify the appropriate $\mathbb{Z}_{L }$ action on the worldsheet fields. To that end, it is useful to define the complex combinations
\begin{equation}\label{eq:cplxbosons}
    Z_4=X^6+i X^7\,,\qquad Z_5=X^8+i X^{9}\,,
\end{equation}
which parameterise two orthogonal (internal) $\mathbb{C}$ planes. The orbifold action we want to quotient by takes the form
\begin{equation}\label{eq:orbifoldaction}
    \Gamma_{L }: \qquad Z_4\to e^{\frac{2\pi i}{L }}Z_4\,,\quad Z_5\to e^{-\frac{2\pi i}{L }}Z_5\,,\quad H_4\to H_4 + \frac{2\pi }{L } \,,\quad H_5\to H_5 - \frac{2\pi}{L } \,.
\end{equation}
Untwisted vertex operators are generated by acting with the projector
\begin{equation}\label{eq:projector}
    \mathcal{P}=\frac{1}{L }\sum_{n=0}^{L -1}\Gamma_{L }^n\,,
\end{equation}
on regular type IIB vertex operators, for example,
\begin{equation}
:e^{i k_6 X^6}:\qquad \rightarrow\qquad \frac{1}{L }\sum_{n=0}^{L -1}:e^{ik_6\cos(\frac{2\pi n}{L })X^6}::e^{-ik_6\sin(\frac{2\pi n}{L })X^7}:\,,
\end{equation}
implying that momentum vectors $\Vec{\nu}=(k^6,k^7,k^8,k^9)^\mathsf{T}$ are identified under actions of $\Gamma_{L }$. We will henceforth split the bosonic momentum vectors into the ``internal part'' $\vec{\nu}$ and the ``external part'' $\vec{\mathbf{k}}=(k^0,k^1,k^2,k^3,k^4,k^5)^\mathsf{T}$ and omit the vector arrows. 

To this set of untwisted vertex operators we have to add the twisted sector, generated by twist operators $\Sigma_n$ which generate the appropriate monodromy
\begin{equation}
    \phi(z + \epsilon e^{2\pi i})\Sigma_n(z)= \Gamma_{L }^n\left(\phi(z+\epsilon)\right)\Sigma_n(z)\,,
\end{equation}
for an arbitrary field $\phi$. For the fermionic fields, this monodromy can be created by a vertex operator but for the bosonic fields the twist operator $\sigma_n$ may only be defined implicitly by
\begin{equation}\label{eq:twistevertexop}
    \Sigma_n(z)=:e^{i \frac{n}{L } H_4(z)+i\frac{L -n}{L }H_5(z)}:\sigma_n(z)\,.
\end{equation}
A completion by anti-holomorphic parts is implied. The conformal weight of $\sigma_n$ is known to be $h(\sigma_n)=\bar h(\sigma_n)=\tfrac{n(L -n)}{L ^2}$ \cite{Dixon:1986qv}, which, together with the fermionic vertex operator results in a total twist operator contribution of $h(\Sigma_n)=\bar h(\Sigma_n)=\frac{1}{2}$\,. This is precisely enough to raise the twisted sector vacuum to the massless level, once ghosts and superghosts are accounted for. Note, however, that we cannot combine $\sigma_n$ with an internal momentum operator $:e^{i\nu\cdot X}:$.\footnote{Performing the OPE of these operators returns the twist operator $\sigma_n$ and possible fractional modes excited on top, but crucially, the vertex operator is annihilated in the process.} This confirms our expectation that twisted sector states are localised on the 6d fixed subspace of the orbifold action, in which it may move with momentum $\mathbf{k}$.

The appearance of the bosonic twist operators $\sigma_n$ makes explicit calculations slightly more cumbersome. A standard prescription to deal with such operators is the introduction of a covering map $z(t)$ \cite{Hamidi:1986vh,Dixon:1986qv}, which locally around the twist operators behaves as $ z(t)\sim t^{L }$. This allows us to patch together the bosonic monodromy to make the bosons periodic in $t$. In this covering space, one then performs ordinary free boson calculations without any twist operator insertions and pulls back the result at the very end. For our purposes it will be enough to restrict our attention to the simplest twisted amplitude available, involving only two twisted vertex operators. We use conformal symmetry to move them to $0$ and $z_\infty$, respectively, which results in the simple covering map 
\begin{equation}\label{eq: covering map}
    z(t)=\frac{t^L  z_\infty }{t^L +z_\infty}\qquad \underset{z_\infty\to\infty}{\longrightarrow}\qquad z(t)=t^{L }\,.
\end{equation}
One can easily check that an expansion around $t=0$ and $t=\infty$ yields the appropriate monodromies.

We have now set up most of the theory, aside from a proper discussion of  ghost-, superghost- and BRST-properties. As this discussion is parallel to the case of usual type IIB strings on flat space, we refer to the textbooks, e.g. \cite{Blumenhagen:2013fgp}. We now want to compute a simple tree-level four-point amplitude involving two twisted and two untwisted states and analyse how it differs from the Virasoro-Shapiro amplitude we know from type IIB string theory on flat space \cite{Gross:1986iv}. Since we are mostly interested in the overall coefficient, we will not work out the full tensor structure of twisted amplitudes, but keep a full analysis of the precise kinematics as an interesting challenge for future research. Without further ado, let us consider the following NS-NS sector vertex operators
\begin{align}
    \mathcal{V}_0(z_0,\bar{z}_0)&=\,:c(z_0)\tilde{c}(\bar{z}_0) e^{-\phi(z_0)}e^{-\tilde{\phi}(\bar{z}_0)}\Sigma_n e^{i \mathbf{k}_0 \cdot X(z_0,\bar{z}_0)}:\,,\label{eq: Vertex0}\\
    \mathcal{V}_\infty(z_\infty,\bar{z}_\infty)&=\,:c(z_\infty)\tilde{c}(\bar{z}_\infty) e^{-\phi(z_\infty)}e^{-\tilde{\phi}(\bar{z}_\infty)}\Sigma_{L -n} e^{i \mathbf{k}_\infty \cdot X(z_\infty,\bar{z}_\infty)}:\,,\label{eq: Vertexinf}\\
    \mathcal{V}_1(z_1,\bar{z}_1)&=\,:c(z_1)\tilde{c}(\bar{z}_1) e^{-\phi(z_1)}e^{-\tilde{\phi}(\bar{z}_1)}e^{iH_3(z_1,\bar{z_1})}e^{i \mathbf{k}_1 \cdot X(z_1,\bar{z}_1)}\mathcal{P}e^{i \nu\cdot X(z_1,\bar{z_1})}:\,,\label{eq: Vertex1}\\
    \mathcal{V}_z(z,\bar{z})&=\,:e^{-\phi(z)}e^{-\tilde{\phi}(\bar{z})}e^{iH_3(z,\bar{z})}e^{i \mathbf{k}_z \cdot X(z,\bar{z})}\mathcal{P}e^{-i \nu\cdot X(z,\bar{z})}:\,.\label{eq: Vertexz}
\end{align}
Here, $c,\bar{c}$ are $bc$-ghost insertions in anticipation of fixing $z_0\to0$, $z_\infty\to\infty$ and $z_1\to 1$ by way of $\mathrm{PSL}(2,\mathbb{C})$ transformations on the Riemann sphere. The final integration is only over the position of $\mathcal{V}_z(z,\bar{z})$. The fields $\phi, \bar{\phi}$ denote the bosonisation of the $\beta\gamma$-superghosts. They signal that all these operators are in the $(-1,-1)$-picture. This is actually a problem since the overall picture number on the sphere should be $(-2,-2)$. To alleviate the mismatch, we have to raise the picture of two of the vertex operators by acting with the picture raising operator (PRO)
\begin{equation}
    \mathrm{PRO}\sim2 e^\phi T_{\text{F}}+\dots\,,\qquad T_{\text{F}}\sim \frac{i}{2}\psi_\mu \partial X^\mu+\dots\,.
\end{equation}
Although this procedure would be difficult to perform in general, we note that the polarisations of the operators in \eqref{eq: Vertex1} and \eqref{eq: Vertexz} have been chosen such that the contraction of the $H_3$ vertex operators vanishes unless picture raising generates precisely the necessary operators to balance out the $H_3$-charge. We can therefore remain ignorant of the full expressions and simply extract the required terms
\begin{align}
    [\mathrm{PRO}\,,\mathcal{V}_0(z_0,\bar{z}_0)]&\supset\,:c(z_0)\tilde{c}(\bar{z}_0) e^{-iH_3(z_0,\bar{z}_0)}\Sigma_n e^{i \mathbf{k}_0 \cdot X(z_0,\bar{z}_0)}:\,,\\
    [\mathrm{PRO}\,,\mathcal{V}_\infty(z_\infty,\bar{z}_\infty)]&\supset\,:c(z_\infty)\tilde{c}(\bar{z}_\infty) e^{-iH_3(z_\infty,\bar{z}_\infty)}\Sigma_{L -n} e^{i \mathbf{k}_\infty \cdot X(z_\infty,\bar{z}_\infty)}:\,,
\end{align}
where we dropped a kinematic prefactor and assumed that there is some generic momentum in the $(4,5)$-plane for both vertex operators. We may equally consider any other orientation of the untwisted operators, so this assumption is merely a kinematic one and not important to our argument. 

We can now get to work on computing the four-point function
\begin{equation}
    \int_{\mathbb{C}}\dd z\dd \bar{z}\,\langle\mathcal{V}_\infty(z_\infty,\bar{z}_\infty)\mathcal{V}_z(z,\bar{z})\mathcal{V}_1(z_1,\bar{z}_1)\mathcal{V}_0(z_0,\bar{z}_0)\rangle\,.
\end{equation}
Up to inconsequential numerical factors, most contributions can be computed directly. We split the bosonic and fermionic CFTs into ``external'' and ``internal'' parts probing $\mathbb{R}^{1,5}$ and $\mathbb{C}^2/\mathbb{Z}_{L }$, respectively. Sending $z_0\to 0$ and $z_1\to1$, but keeping $z_\infty$ finite for now, we end up with the contributions
\begin{equation}\begin{split}
    \mathcal{C}^{B}_{\text{ext}}
    &\sim\abs{z_\infty-z}^{\alpha' \mathbf{k}_\infty\cdot \mathbf{k}_z}\abs{z_\infty-1}^{\alpha' \mathbf{k}_\infty\cdot \mathbf{k}_1}\abs{z_\infty}^{\alpha' \mathbf{k}_\infty\cdot \mathbf{k}_0}\abs{z}^{\alpha' \mathbf{k}_z\cdot \mathbf{k}_0}\abs{z-1}^{\alpha' \mathbf{k}_z\cdot \mathbf{k}_1}\,,\\
    \mathcal{C}^{F}_{\text{ext}}
    &\sim\abs{z_\infty-z}^{-2}\abs{z_\infty-1}^{-2}\abs{z_\infty}^2\abs{z}^{-2}\abs{z-1}^2\,,\\
    \mathcal{C}^{F}_{\text{int}}&\sim\abs{z_\infty}^{-2 +\frac{4n(L -n)}{L ^2}}\,,\\
    \mathcal{C}^{bc}&\sim\abs{z_\infty}^2\abs{z_\infty-1}^2\,,\\
    \mathcal{C}^{\beta\gamma}&\sim\abs{z-1}^{-2}\,.
\end{split}\label{eq: contrest}\end{equation}
The only remaining contribution from internal bosons and twist operators
\begin{equation}
    \mathcal{C}^{B}_{\text{int}}(z,\bar{z})=\langle\sigma_{L -n}(z_\infty,\bar{z}_\infty)\mathcal{P}e^{-i \nu\cdot X(z,\bar{z})}\mathcal{P}e^{i \nu\cdot X(1,1)}\sigma_{n}(z_0,\bar{z}_0)\rangle\,,
\end{equation}
requires a treatment with covering space methods. Note first that overall momentum conservation restricts the momenta $\nu$ and $-\nu$ to sum to $0$. The two projection operators $\mathcal{P}$ therefore cannot generate non-vanishing cross-terms and their only contributions are simultaneous rotations of both $\nu$ and $-\nu$ by an action of $\Gamma_L ^n$, which we shall denote by $\nu_n$ and $-\nu_n$. The correlator for each of these rotations is identical so we may drop the projection operators completely in favour of a single factor $L ^{-1}$. When employing the covering map \eqref{eq: covering map}, the primary operators involved have to be transformed according to 
\begin{equation}
    \mathcal{O}(z,\bar{z})=\abs{z'(t)}^{-\Delta}\mathcal{O}(t,\bar{t})\,.
\end{equation}
We will expand this prefactor in large $z_\infty$. Furthermore, the untwisted operators inserted at $1$ and $z$ have multiple pre-images in covering space, so they decompose into $L $ mirror operators at locations $\xi^{m}_{L }$ and $\sqrt[k]{z}\xi^{m}_{L }$\,, where $\xi_{L }=e^{\frac{2\pi i }{L }}$ denotes the $L ^{th}$ root of unity and $m\in\{0,1,...L-1\}$. We thus end up with a covering map expression of the form
\begin{equation}\begin{split}
    \mathcal{C}^{B}_{\text{int}}=L ^{-1-\alpha'\nu^2-4\frac{n(L -n)}{L ^2}}&\abs{t}^{-\frac{(L -1)}{2}\alpha'\nu^2}\abs{z_\infty}^{-4\frac{n(L -n)}{L ^2}}\abs{\epsilon}^{4\frac{n(L -n)}{L }}\times\\
    & \langle \prod_{m=0}^{L -1}:e^{i \nu_{nm} \cdot X(\xi^{m}_{L },\xi^{-m}_{L })}:\prod_{l=0}^{L -1}:e^{i \nu_{nl} \cdot X(t\xi^{l}_{L },\bar t\xi^{-l}_{L })}:\rangle^{\frac{1}{L }}\,.
\end{split}\end{equation}
We used a cut-off $\epsilon$ around the twist-operator insertions at $t=0$ and $t=\infty$. For a precise treatment of the twist operator regularisation see a parallel discussion in \cite{Lunin:2000yv}. Essentially, these factors may be absorbed in the normalisation of the twist-operators. 

The various mirror images of the individual vertex operators may contract either among themselves, resulting in the expressions ($n\neq0$)
\begin{align}
    \prod_{l<m}\vert\xi^{l}_{L }-\xi^{m}_{L }\vert^{\alpha' \nu^2 \cos \left(\frac{2\pi n}{L }(l-m)\right)}&=\prod_{m=1}^{L -1}\abs{1-\xi^{m}_{L }}^{\frac{L }{2}\alpha' \nu^2 \cos \left(\frac{2\pi n}{L }m\right)}\,,\\
    \prod_{l<m}\vert t(\xi^{l}_{L }-\xi^{m}_{L })\vert^{\alpha' \nu^2 \cos \left(\frac{2\pi n}{L }(l-m)\right)}&=\abs{t}^{-\frac{L }{2}\alpha'\nu^2}\prod_{m=1}^{L -1}\abs{1-\xi^{m}_{L }}^{\frac{L }{2}\alpha' \nu^2 \cos \left(\frac{2\pi n}{L }m\right)}\,,
\end{align}
or with mirrors of the other respective vertex operator, resulting in the contribution
\begin{equation}
    \prod_{l,m=0}^{L -1}\abs{\xi_{L }^l-t \xi_{L }^m}^{\alpha' \nu^2 \cos \left(\frac{2\pi n}{L }(l-m)\right)}=\prod_{m=0}^{L -1}\abs{1-t\xi^{m}_{L }}^{-L \alpha' \nu^2 \cos \left(\frac{2\pi n}{L }m\right)}\,.
\end{equation}
Combining the various contributions and returning to the original worldsheet coordinates, we end up with
\begin{equation}\begin{split}\label{eq: contbos}
    \mathcal{C}^{B}_{\text{int}}=\frac{C}{L}\abs{z}^{-\frac{\alpha'\nu^2}{2}}\abs{z_\infty}^{-4\frac{n(L -n)}{L ^2}} \sum_{l=0}^{L-1}\prod_{m=0}^{L -1}\abs{1-\sqrt[L ]{z}\xi^{(m+l)}_{L }}^{-\alpha' \nu^2 \cos \left(\frac{2\pi n}{L }m\right)}\,,
\end{split}\end{equation}
where we abbreviated a constant factor
\begin{equation}
    C=L ^{-1-\alpha'\nu^2}\prod_{m=1}^{L -1}\left\vert1-\xi^{m}_{L }\right\vert^{\alpha' \nu^2 \cos \left(\frac{2\pi n}{L }m\right)}\,.
\end{equation}
Note that we restricted ourselves to the principal branch of $\sqrt[L]{z}$ and accounted for this choice by averaging over all preimages. 

We may now collect the factors \eqref{eq: contrest} and \eqref{eq: contbos} and observe in particular that all terms involving $z_\infty$ nicely cancel in the $z_\infty\to\infty$ limit, once the kinematic conditions $\mathbf{k}_0+\mathbf{k}_1+\mathbf{k}_\infty+\mathbf{k}_z=0$ and $\mathbf{k}_\infty^2=0$ are enforced. To make contact with the usual notation, we define six-dimensional Mandelstam variables
\begin{equation}\label{eq:Mandelstam1}
s=-(\textbf{k}_{0}+\textbf{k}_{1})^{2}\,,\quad t=-(\textbf{k}_{0}+\textbf{k}_{\infty})^{2}\,,\quad u=-(\textbf{k}_{0}+\textbf{k}_{z})^{2}\,.
\end{equation}
Since we consider massless external states, the 6d momenta satisfy
\begin{equation}\label{eq:Mandelstam2}
\textbf{k}_{0}^{2}=\textbf{k}_{\infty}^{2}=0\,,\quad\textbf{k}_{1}^{2}=\textbf{k}_{z}^{2}=-\nu^{2}\qquad \Rightarrow\qquad s+t+u=2\nu^2\,.
\end{equation}
The final integral expression now takes the form
\begin{equation}\label{eq:TwistedVSintegral}
\int_{\mathbb{C}} \dd z \dd \bar{z}\, \abs{z}^{-\frac{\alpha'u}{2}-2}\abs{1-z}^{-\frac{\alpha'}{2}(t-2\nu^2)}\frac{1}{L}\sum_{l=0}^{L-1}\prod_{m=0}^{L -1}\abs{1-\sqrt[L ]{z}\xi^{(m+l)}_{L }}^{-\alpha' \nu^2 \cos \left(\frac{2\pi n}{L }m\right)}\,,
\end{equation}
which is generally hard to evaluate. 
\subsection{Low-energy expansion of the integral}
Since we do not expect closed-form expressions for the integral \eqref{eq:TwistedVSintegral}, obtaining its low-energy expansion requires analysing its pole structure directly at the level of the integrand. We first study the poles in the $u$-channel, in which virtual twisted states are exchanged. These poles are associated to a potential $\vert z\vert\to 0$ divergence of \eqref{eq:TwistedVSintegral}, where the vertex operator inserted at $z$ approaches the one inserted at $0$.

The main divergent term in this limit is $\abs{z}^{-\frac{\alpha'u}{2}-2}$, while the other terms remain finite. We may therefore consider a small-$\alpha'$ expansion of these additional terms. For the $\vert 1-z\vert$ factor in \eqref{eq:TwistedVSintegral}, we expand as \begin{equation}\begin{split}\label{eq:binompiece}
\vert 1-z\vert^{-\tfrac{\alpha'(t-2\nu^2)}{2}}=&1+\frac{\alpha'(t-2\nu^2)}{4}\sum_{r=1}^{\infty}\frac{z^r+\bar{z}^r}{r}\\
&+\frac{\alpha'^2(t-2\nu^2)^2}{32}\sum_{r,s\geq 1}\frac{(z^r\bar{z}^s+z^s\bar{z}^r)+(z^{r+s}+\bar{z}^{r+s})}{rs}+\order{\alpha'^3}\,.
\end{split}\end{equation} To deal with the product factor, we use the fact that for $\vert z\vert<1$
\begin{equation}
\log\left\vert 1-\sqrt[L]{z}\xi_{L }^{(m+l)}\right\vert=\Re\left[\log(1-\sqrt[L]{z}\xi_{L }^{(m+l)})\right]=-\frac{1}{2}\sum_{k= 1}^{\infty}\frac{z^{\tfrac{k}{L}}\xi_{L }^{(m+l)k}+\overline{z}^{\tfrac{k}{L}}\xi_{L }^{-(m+l)k}}{k}\,.
\end{equation}
The logarithm of the product factor in \eqref{eq:TwistedVSintegral} may then be written as \footnote{To do so, we note that only terms with $k=\pm n\,\mathrm{mod}\,L $ are present in the final sum. This is a consequence of evaluating sums of the form $$\sum_{m=0}^{L -1}\cos\left(\frac{2\pi n}{L }m\right)\xi_{L }^{(m+l)k}=\frac{L \xi_{L}^{lk}}{2}\left(\delta_{k,n\,\mathrm{mod}\,L }+\delta_{k,-n\,\mathrm{mod}\,L }\right)\,.$$ We may thus parametrise the surviving terms as $k=pL +n$ and $k=(p+1)L-n$ where $p\in\mathbb{Z}_{\geq 0}$.}
\begin{equation}\begin{split}\label{eq:Lerch}
\log\Bigg[\prod_{m=0}^{L -1}\vert1-\sqrt[L]{z}\xi_{L }^{(m+l)}&\vert^{-\alpha'\nu^{2}\cos\left(\tfrac{2\pi n}{L }m\right)}\Bigg]=\frac{\alpha'\nu^{2}}{2}\sum_{p=0}^{\infty}\mathrm{Re}\left[\frac{\xi_{L}^{nl}z^{p +\tfrac{n}{L}}}{p +\tfrac{n}{L}}+\frac{\xi_{L}^{-nl}z^{(p+1) -\tfrac{n}{L}}}{(p+1) -\tfrac{n}{L}}\right]\\
&\;\quad=\frac{\alpha'\nu^{2}}{2}\mathrm{Re}\bigg[\xi_{L}^{ln}z^{\tfrac{n}{L}}\Phi\left(z,1,\frac{n}{L }\right)+\xi_{L}^{-nl}z^{1 -\tfrac{n}{L}}\Phi\left(z,1,1-\frac{n}{L }\right)\bigg]\,,
\end{split}\end{equation} 
making manifest its representation in terms of  the Lerch transcendent $\Phi(z,s,\alpha)$. Exponentiating the above and performing a small $\alpha'$ expansion yields
\begin{equation}\label{eq:twistedpiece}\begin{split}
\frac{1}{L}\sum_{l=0}^{L-1}\prod_{m=0}^{L -1}\left\vert1-\sqrt[L]{z}\xi_{L }^{m+l}\right\vert&^{-\alpha'\nu^{2}\cos\left(\tfrac{2\pi n}{L }m\right)}=\frac{1}{L}\sum_{l=0}^{L-1}\Bigg[1+\frac{\alpha'\nu^2}{4}\sum_{q\in\mathcal{Q}}\frac{\xi_{L}^{\sigma nl}z^{q}+\xi_{L}^{-\sigma nl}\bar{z}^{q}}{q}+\\
&\frac{\alpha'^2\nu^4}{32}\sum_{q,q'\in\mathcal{Q}}\frac{(\xi_{L}^{\sigma nl}z^q+\xi_{L}^{-\sigma nl}\bar{z}^q)(\xi_{L}^{\sigma' nl}z^{q'}+\xi_{L}^{-\sigma' nl}\bar{z}^{q'})}{qq'}+\order{\alpha'^3}\Bigg]\,,
\end{split}\end{equation}
where we have introduced the indices $q,q'\in\mathcal{Q}:=\{p +\tfrac{n}{L},\,p+1 -\tfrac{n}{L}:\,p\in\mathbb{Z}_{\geq 0}\}$ and abbreviated the sign factors
\begin{equation}
    \sigma=\sigma(q)=\begin{cases}
    +1\,,\;\mathrm{if}\;q=p+\frac{n}{L}\\
    -1\,,\;\mathrm{if}\;q=p+1-\frac{n}{L}\,,
    \end{cases}\qquad \sigma'=\sigma(q')\,.
\end{equation}
 Exchanging the order of summation to first perform the sum over $l$, the linear term in \eqref{eq:twistedpiece} vanishes automatically since $\sum_{l=0}^{L-1}\xi_{L}^{\pm nl}=0$ for $n\neq 0$. The additional phases $l$ arising from the averaging over the different covering maps (cf. \eqref{eq: contbos}) enforce a set of selection rules on the quadratic term due to the fact that
\begin{equation}
    \sum_{l=0}^{L-1}\xi_{L}^{ln(\sigma \pm \sigma')}=L \delta_{\sigma,\mp\sigma'}\,.
\end{equation}
The above condition, in turn, implies that the only surviving terms must be of one of two forms: either $z^q\bar{z}^{q'}$ terms with $\sigma=\sigma'$  or (anti)holomorphic terms $z^{q+q'}$ ($\bar{z}^{q+q'})$ with $\sigma=-\sigma'$, from which we can deduce that $q+q'\in\mathbb{Z}_{\geq 1}$. One finds that \eqref{eq:twistedpiece} reduces to
\begin{equation}
1+\frac{\alpha'^2\nu^4}{32}\sum_{q,q'\in\mathcal{Q}}\frac{(z^{q}\bar{z}^{q'}+z^{q'}\bar{z}^q)\delta_{\sigma,\sigma'}+(z^{q+q'}+\bar{z}^{q+q'})\delta_{\sigma,-\sigma'}}{qq'}+\order{\alpha'^3}\,.
\end{equation}
Collecting these expansions, the integrand of \eqref{eq:TwistedVSintegral} may be written as
\begin{equation}\begin{split}\label{eq:smalla'exp}
\vert z \vert^{-2-\tfrac{\alpha' u}{2}}\Bigg[&1+\frac{\alpha'(t-2\nu^2)}{4}\sum_{r=1}^{\infty}\frac{z^r+\bar{z}^r}{r}\\
&+\frac{\alpha'^2(t-2\nu^2)^2}{32}\sum_{r,s\geq 1}\frac{(z^r\bar{z}^s+z^s\bar{z}^r)+(z^{r+s}+\bar{z}^{r+s})}{rs}\\
&+\frac{\alpha'^2\nu^4}{32}\sum_{q,q'\in\mathcal{Q}}\frac{(z^{q}\bar{z}^{q'}+z^{q'}\bar{z}^q)\delta_{\sigma,\sigma'}+(z^{q+q'}+\bar{z}^{q+q'})\delta_{\sigma,-\sigma'}}{qq'}+\order{\alpha'^3}\Bigg]\,,
\end{split}\end{equation}
in a small $\alpha'$ expansion. We now transform to polar coordinates on $\mathbb{C}$ defined by $\rho\in[0,\rho_{0}]$ and $\theta\in[0,2\pi)$, where $\rho_{0}$ is an arbitrary cut-off scale that we can set to 1. Upon performing the angular integral, both the linear term and the (anti-)holomorphic quadratic term coming from \eqref{eq:binompiece} vanish. The same is true for the (anti-)holomorphic terms from \eqref{eq:twistedpiece} since the selection $\delta_{\sigma,-\sigma'}$ rule ensures that $q+q'\in\mathbb{Z}_{\geq 1}$. The angular integral further projects the mixed terms of the form $z^r\bar{z}^s$ down to the diagonal part $r=s$ of the sum. The selection rule $\delta_{\sigma,\sigma'}$ ensures that $q-q'\in\mathbb{Z}$, which allows us to use the same argument for the mixed terms $z^q\bar{z}^{q'}$. Performing the radial integral on the surviving terms yields \footnote{We are regularising the integral by shifting $u$ into the convergent region. The divergence of the integral then generates poles in $u$. }
\begin{equation}\label{eq:uradialint}
\frac{2\pi}{-\tfrac{\alpha'u}{2}}+\frac{\pi \alpha'^2(t-2\nu^2)^2}{8}\sum_{r\geq 1}\frac{1}{r^2(2r-\tfrac{\alpha'u}{2})}+\frac{\pi\alpha'^2\nu^4}{8}\sum_{q\in\mathcal{Q}}\frac{1}{q^2(2q-\tfrac{\alpha'u}{2})}+\order{\alpha'^3}\,.
\end{equation}
From the above expression, we obtain poles in the $u$-channel whenever
\begin{equation}\label{eq:upoles}
    u=\frac{4}{\alpha'}r\,,\quad u=\frac{4}{\alpha'}\left(p+\frac{n}{L}\right)\,,\quad u=\frac{4}{\alpha'}\left(p+1-\frac{n}{L}\right)\,,
\end{equation}
where $r,p\in\mathbb{Z}_{\geq 0}$. This is in agreement with the masses of excitations found in Appendix \ref{App: Spectrum}. In particular, for $L=1$, we reproduce exactly the poles of the usual Virasoro-Shapiro amplitude at $u=\frac{4}{\alpha'}r$ for integer $r$.

For small $\alpha'$, the leading pole at $u=0$ dominates \eqref{eq:uradialint} and we may therefore expand \eqref{eq:uradialint} in $\alpha'$ to find 
\begin{equation}\label{eq:unearpole}
-\frac{4\pi}{\alpha' u}+\frac{\pi\alpha'^2(t-2\nu^2)^2}{16}\sum_{r\geq 1}\frac{1}{r^3}+\frac{\pi\alpha'^2\nu^4}{16}\sum_{q\in\mathcal{Q}}\frac{1}{q^3}+\order{\alpha'^3}\,.
\end{equation}

We now turn our attention to the $s$-channel divergences of \eqref{eq:TwistedVSintegral}, corresponding to the limit where the vertex operators inserted at 0 and 1 approach each other. These divergences become manifest in the regime where $\vert z\vert\to\infty$. It is therefore convenient to transform $z\to \frac{1}{w}$, such that the divergent limit is $\vert w\vert\to0$. The integral \eqref{eq:TwistedVSintegral} becomes
\begin{equation}\label{eq:wintegral}
\int\dd w\dd\bar{w}\,\vert w\vert ^{-\tfrac{\alpha's}{2}-2}\vert 1-w\vert^{-\tfrac{\alpha'(t-2\nu^2)}{2}}\,\frac{1}{L}\sum_{l=0}^{L-1}\prod_{m=0}^{L -1}\left\vert 1-\sqrt[L]{w}\xi_{L }^{(m+l)}\right\vert^{-\alpha'\nu^2\cos\left(\tfrac{2\pi n}{L }m\right)}\,,
\end{equation}
which has the same form as \eqref{eq:TwistedVSintegral} with $u\leftrightarrow s$ exchanged.
The $s$-channel poles thus appear at the same locations \eqref{eq:upoles} as in the $u$-channel and the $\alpha'$-expansion singles out the leading pole at $s=0$, resulting in the expression
\begin{equation}\label{eq:snearpole}
  -\frac{4\pi}{\alpha's}+\frac{\pi\alpha'^2(t-2\nu^2)^2}{16}\sum_{r\geq 1}\frac{1}{r^3}+\frac{\pi\alpha'^2\nu^4}{16}\sum_{q\in\mathcal{Q}}\frac{1}{q^3}+\order{\alpha'^3}\,,
\end{equation}
in complete analogy with the $u$-channel. 

We finally observe that there is no $t$-channel divergence at small $\alpha'$, since the pole at $t=2\nu^2$ is absent. This is due to the polarisations chosen in \eqref{eq: Vertex1} and \eqref{eq: Vertexz}, which generate a kinematic prefactor to the amplitude that scales as $(t-2\nu^2)^2$, suppressing the pole at $t=2\nu^2$. The Taylor expansions around $z=0$ and $w=0$ may be trusted all the way to $\abs{z}=\abs{w}=1-\epsilon$. Since the integral is bounded on the annulus $\vert z\vert\in(1-\varepsilon,1+\varepsilon)$, the contribution from this region may be dropped in the limit $\varepsilon\to 0$. We may thus simply add \eqref{eq:unearpole} and \eqref{eq:snearpole} together to obtain the final expression for the small $\alpha'$ expansion of \eqref{eq:TwistedVSintegral}, which reads
\begin{equation}\label{eq:thelowenergyexp}
\frac{4\pi}{\alpha'}\frac{(t-2\nu^2)}{su}+\frac{\pi\alpha'^2}{16}\Bigg[2(t-2\nu^2)^2\zeta(3)-\nu^4\bigg[\psi^{(2)}\left(\frac{n}{L}\right)+\psi^{(2)}\left(1-\frac{n}{L}\right)\bigg]\Bigg]+\order{\alpha'^3}\,.
\end{equation}
The above shows the presence of a correction term originating from twisted virtual particles that generates precisely the combination of polygamma functions observed in \eqref{eq: Zl-case}. We note that its kinetic factor does not include a dependence on the 6d Mandelstams $(s,t,u)$, since it is generated purely by internal modes. Untwisted operators of the form \eqref{eq: Vertex1} without internal momentum ($\nu=0$) would not generate polygamma terms in the amplitude. This can also be seen directly from the integral form \eqref{eq:TwistedVSintegral}, which reduces to the conventional Virasoro-Shapiro amplitude (restricted to 6d) when $\nu=0$. 

One could of course also consider operators with excitations in the internal dimensions. We expect that they result in the same non-trivial transcendental structure as \eqref{eq:thelowenergyexp} but with more involved kinematic factors. Similarly, an analysis on BMN-type backgrounds should only change the kinematic structure of the low-energy expansion and thus reproduce \eqref{eq:BMNexpansion2ptfunct}. We would like to explore further twisted correlators in the future, but take the current calculation as a successful proof of concept suggesting that the polygamma factors in \eqref{eq: Zl-case} should be matched to $(\alpha')^3$-corrections that arise from the low-energy expansion of twisted Virasoro-Shapiro amplitudes.

\section{Conclusion}\label{sec: 4}

In this paper, we investigated type IIB string theory on $\mathrm{AdS}_5\times S^5/\mathbb{Z}_L$ orbifold backgrounds. In particular, we have studied to what extent the subleading terms in the strong-coupling expansion of twisted correlators \eqref{eq: Zl-case} can be matched to $\alpha'$-corrections of the effective 6d supergravity theory for twisted string modes. We presented two possible approaches to this question.

The first approach, outlined in Section \ref{sec: 2}, was motivated by the resolution procedure $\mathbb{R}^{1,5}\times \mathbb{C}^2/\mathbb{Z}_L\to \mathbb{R}^{1,5}\times \textrm{GH}_L$, from which one may deduce an effective 6d supergravity action for the twisted sector fields. Although this approach can in principle be applied to curved backgrounds such as $\mathrm{AdS}_5\times S^5/\mathbb{Z}_L$ \cite{Gukov:1998kk}, a full solution of the equation of motion that resolves the orbifold singularity is quite difficult to construct. Even in the simplest case of $L=2$, only an irregular geometry is known \cite{Skrzypek:2023fkr}. However, at least in the pp-wave limit, such geometries can be constructed and an effective 6d theory may be deduced. 

We then attempted to apply this logic at the level of $(\alpha')^3$-corrections. The full expression for the $(\alpha')^3$-term in 10d is so far unknown, so we can only argue for the presence of certain appropriate covariant tensor structures, which we can tune precisely in order to match the kinetic factors in \eqref{eq: Zl-case}. However, any such corrections necessarily feature a factor $\zeta(3)$, which is universal to the 10d $(\alpha')^3$-correction term. This is in conflict with the localisation result \eqref{eq: Zl-case}.

Motivated by this observation, we turned to the second approach in Section \ref{sec: 3}, where we considered the $(\alpha')^3$-terms in the low-energy expansion of a sample string amplitude involving twisted sector states. In contrast to the usual Virasoro-Shapiro amplitude, we now have to insert twisted vertex operators for these twisted external states. As a consequence, twisted states also appear as resonances, changing the pole structure of the amplitude. We showed that this results in the appearance of twist-dependent polygamma factors alongside the usual $\zeta(3)$, in agreement with the localisation result \eqref{eq: Zl-case}. 

This result should be taken as a proof of concept that the polygamma factors are a consequence of the exchange of virtual twisted states and thus universal to twisted sector amplitudes. A more thorough analysis of string amplitudes in the various sectors and polarisations would be an interesting challenge for future research. This would give us direct access to the effective $(\alpha')^3$-corrections and their tensor structure in 6d.
However, these string amplitudes can only be constructed in the locally flat $\mathbb{R}^{1,5}\times \mathbb{C}^2/\mathbb{Z}_L$ space, and not in $\mathrm{AdS}_5\times S^5/\mathbb{Z}_L$, since quantisation of the string on RR-backgrounds is beyond reach. One may try to circumvent this issue by applying the AdS Virasoro-Shapiro Ansatz introduced in \cite{Alday:2023mvu}. Aside from the pole structure and residues of the flat space orbifold amplitude, this would require working out the specific details of the (super)conformal block expansion in the dual quiver gauge theory and perhaps some input from integrability. We would like to investigate this direction in the near future.

Aside from the special cases of $L=2,3,4,6$, where the polygammas can be expressed in terms of $\zeta(3)$, our results seem to imply that the resolution procedure investigated in Section \ref{sec: 2} is incomplete and does not capture string amplitudes with twisted sector external and intermediate states. This statement is puzzling given that string theory on a completely smooth resolution space $\mathrm{AdS}_5\times \mathcal{M}_L^5$ should have a sensible low-energy expansion, which is expected to take the form of the string effective action (supergravity plus additional $\alpha'$-correction terms) evaluated on $\mathrm{AdS}_5\times \mathcal{M}_L^5$ background. This reasoning could break down at several stages. The most disappointing scenario would be the inexistence of a globally well-defined resolution $\mathrm{AdS}_5\times \mathcal{M}_L^5$. It would then be interesting to clearly isolate the obstruction, given that approximate arguments seem successful at supergravity level. A more likely scenario is that the approximations we relied on, especially when approximating field configurations close to the singularity were too naive. There is also a good chance that taking the resolution size to 0 we should have employed a more subtle renormalisation procedure. Still, it seems rather challenging to explain the appearance of polygamma functions instead of the generic $\zeta(3)$ with purely geometric arguments. 

To settle these questions satisfactorily, one may require a complete study of string theory on the resolution space $\mathrm{AdS}_5\times \mathcal{M}_L^5$, which is far outside our current technical capabilities. In fact, the historical motivation for studying orbifolds \cite{Dixon:1985jw,Dixon:1986jc} was to probe the smooth but complicated geometry of the K3-manifold by going to a more tractable orbifold limit. The resolution procedure goes against this logic, resulting in an arguably more complicated background from a string-theoretic point of view. Trying to argue for $\alpha'$-corrections at the level of the resolution is thus discouraged by our results. It seems at this stage more feasible to study string amplitudes directly on the locally flat orbifold space and then perhaps evaluate the resulting effective 6d theory on $\mathrm{AdS}_5\times S^1$ background. We presented the first promising steps in this direction. 
\vfill
\section*{Acknowledgments}
We would like to thank to Marco Bill\`o, Sebastian Harris, Grisha Korchemsky, Dennis le Plat, Evgeny Sobko and Alessandro Testa for helpful comments and discussions on related topics. We would especially like to thank Arkady A. Tseytlin for many helpful discussions, collaboration at early stages of this project and for carefully reading the manuscript. CBM acknowledges support from the STFC DTP research studentship grant ST/Y509231/1. TS would like to thank the participants of the \textit{Workshop on Higher-d Integrability} in Favignana (Italy) in 2025 for stimulating discussions that furthered this project. TS would also like to thank the Cluster of Excellence EXC 2121 Quantum Universe 390833306 and the Collaborative Research
Center SFB1624 for creating a productive research environment at DESY.

\newpage
\appendix

\section{Spectrum of type IIB string theory on $\mathbb{R}^{1,5}\times \mathbb{C}^2/\mathbb{Z}_L$}\label{App: Spectrum}
In this appendix we review the construction of the orbifold spectrum from basic mode expansions. We will show that the twisted sectors generate intermediate mass levels in between the massless and the first excited level ($m^{2}=\tfrac{4}{\alpha'}$) of the unorbifolded type IIB string. In the limit $L\to\infty$, this mass gap is densely filled by a continuum of massive twisted states, which we comment on in Appendix \ref{app: largeL}.

We choose to work in light-cone gauge for simplicity. In the RNS formalism, the appropriate supersymmetric worldsheet action 
\begin{equation}
    \mathcal{S}_{\text{RNS, LC}}=-\frac{1}{2\pi}\int \dd z\dd \bar{z}\left(\frac{2}{\alpha'}\partial X^\mu \bar\partial X_\mu+i\psi^\mu \bar\partial\psi_\mu +i \tilde\psi^\mu \partial \tilde\psi_\mu  \right)\,,
\end{equation}
features eight bosons $X^{\mu}(z,\bar{z})$ and eight fermions $\psi^{\mu}(z)$, $\tilde\psi^{\mu}(\bar{z})$ of either chirality. The mode expansions of the left-moving fields are \footnote{Here, the index $\mu\in\{2,3,...,9\}$ denotes the physical directions in light-cone gauge. We will often split $\mu=(a,i)$, where $a=\{2,3,4,5\}$ refer to the physical directions in the external $\mathbb{R}^{1,5}$ space and $i=\{6,7,8,9\}$ correspond to the physical degrees of freedom on the internal $\mathbb{C}^{2}/\mathbb{Z}_{L}$ factor of the background.}
\begin{align}\label{eq:modeexpansions}
    &X^{\mu}(z)=x^{\mu}-i\sqrt{\frac{\alpha'}{2}}\alpha_{0}^{\mu}\ln z+i\sqrt{\frac{\alpha'}{2}}\sum_{m\in\mathbb{Z}\setminus \{0\}}\frac{\alpha^\mu_m}{m}z^{-m}\,,\\
    &\psi^{\mu}(z)=\sum_{r_{\rm{NS/R}}}\psi^{\mu}_{r}\,z^{-r-\tfrac{1}{2}}\,,\qquad r_{\rm{NS}}\in\mathbb{Z}+\tfrac{1}{2}\,,\qquad\;r_{\rm{R}}\in\mathbb{Z}\,,
\end{align}
and similarly for the right-movers $\tilde{X}^{\mu}(z)\,,\,\tilde{\psi}^{\mu}(z)$. Here, we distinguished between Neveu-Schwarz (NS) and Ramond (R) boundary conditions for the fermionic fields. Quantisation promotes the coefficients of the mode expansions to raising and lowering operators, depending on the sign of their mode index. String states are then generated by acting with raising operators on a vacuum. To ensure modular invariance, a GSO projection is required, combining sectors of appropriate boundary conditions.   Before orbifolding, the massless spectrum is generated by th first excited NS state $\psi^{\mu}_{-1/2}\ket{0;k}_{\rm{NS}}$, in the $\textbf{8}_{v}$ of $\mathrm{SO}(8)$, and by one of the massless R-sector ground-states $\ket{\mathrm{R}}^{\alpha}$ or $\ket{\mathrm{R}}^{\dot{\alpha}}$, in the $\mathbf{8}_{s}$ or $\mathbf{8}_{c}$ representations of $\mathrm{SO}(8)$. Each left/right NS/R sector has 8 physical degrees of freedom. Taking the appropriate tensor products therefore results in the familiar $256$ massless states furnishing the supergravity multiplet of 10d type II supergravity.

Upon orbifolding, the untwisted sector is generated by projecting the above 256 states to invariant states under the action \eqref{eq:orbi} of $\Gamma_{L}$. It is convenient to encode the $\Gamma_{L}$ phase of the left-moving fields by an integer charge $q\in\{-1,0,1\}$ 
and to define complex combinations
\begin{equation}
    Z_4=\tfrac{1}{\sqrt{2}}(X^{6}+iX^7)\,,\quad Z_5=\tfrac{1}{\sqrt{2}}(X^8+iX^9)\,,\quad \Psi_4=\tfrac{1}{\sqrt{2}}(\psi^{6}+i\psi^7)\,,\quad \Psi_5=\tfrac{1}{\sqrt{2}}(\psi^8+i\psi^9)\,.
\end{equation}
These have the following charges under \eqref{eq:orbi}
\begin{equation}\begin{split}
    q(Z_4)&=q(\bar{Z}_5)=q(\Psi_4)=q(\bar{\Psi}_5)=1\,,\\ q(Z_5)&=q(\bar{Z}_4)=q(\Psi_{5})=q(\bar{\Psi}_{4})=-1\,.
\end{split}\end{equation}
In addition, we may decompose the R-vacuum into four uncharged vacua and two vacua of charge $\pm1$, respectively. 
Introducing a similar charge, $\tilde{q}$, for the right-moving fields we may impose $\mathbb{Z}_{L}$-invariance as
\begin{equation}
    q+\tilde{q}\equiv 0\;\mathrm{mod}\;L\,.
\end{equation}
For the massless spectrum we have $\abs{q}\leq1$, and if $L>2$ this means that $q=-\tilde{q}$. Thus the only tensor products $q\otimes\tilde{q}$ that survive the orbifold projection combine $\mathbb{Z}_{L}$ charges as $(+1)\otimes(-1),\,0\otimes0$ or $(-1)\otimes(+1)$, resulting in a total of $96$ states in the untwisted sector of the theory.\footnote{For $L=2$, the allowed additional tensor products $(+1)\otimes(+1)$ and $(-1)\otimes(-1)$, bring the total massless states up to $128$, matching the expectation in \cite{Douglas:1996sw}. The 32 `extra' states here combine with the 16 non-local states to form three additional tensor multiplets in the $\bf{3}$ of the broken $\mathrm{SU(2)_{L}}\subset\mathrm{SO}(4)$.} Out of these, 80  organise into the gravity and two tensor multiplets of 6d $\mathcal{N}=(2,0)$ supergravity. The remaining 16 states should be absent from the local spectrum near the orbifold singularity, but form $\mathbb{Z}_{L}$-invariant non-singlet representations of the broken $\mathrm{SU(2)_{L}}$ group. According to \cite{Douglas:1996sw}, these are non-normalisable, untwisted bulk modes on the resolution space which furnish an extra tensor multiplet of 6d $\mathcal{N}=(2,0)$ supergravity.

The $(L-1)$ twisted sectors are generated by imposing the twisted boundary conditions\footnote{Alternatively, they may be seen as generated by the twist vertex operators \eqref{eq:twistevertexop}, which implement the appropriate monodromies.}
\begin{equation}\label{eq:twistedBCsboson}
    Z_4(e^{2\pi i}z)=e^{i\tfrac{2\pi n}{L}}Z_4(z)\,,\qquad Z_5(e^{2\pi i}z)=e^{-i\tfrac{2\pi n}{L}}Z_5(z)\,,
\end{equation}
and similarly for $(\Psi_4,\Psi_5)$ and right-moving fields. Here  $n\in\{1,\dots,L-1\}$ labels the individual twist sectors. These twisted boundary conditions  result in fractional mode numbers at the level of the mode expansions \eqref{eq:modeexpansions}
\begin{align}
    Z_4(z)=i\sqrt{\frac{\alpha'}{2}}\sum_{m\in\mathbb{Z}}\frac{\alpha_{m-\tfrac{n}{L}}^{(4)}}{m-\tfrac{n}{L}}z^{-\left(m-\tfrac{n}{L}\right)}\,,\quad  Z_5(z)=i\sqrt{\frac{\alpha'}{2}}\sum_{m\in\mathbb{Z}}\frac{\alpha_{m+\tfrac{n}{L}}^{(5)}}{m+\tfrac{n}{L}}z^{-\left(m+\tfrac{n}{L}\right)}\,,
\end{align}
with a similar modification for the fermionic mode expansions. Note the absence of zero-modes along the internal directions, implying that twisted modes are localised to the orbifold fixed point. We summarise the moding and minimal contribution to $L_{0}$ (i.e. the corresponding mass gap) of the different oscillators in Table \ref{Tab:modings}.
\begin{table}[htpb]
\centering
\renewcommand{\arraystretch}{1.00}
\setlength{\tabcolsep}{4pt}
\begin{tabular}{|l|cc|cc|}
\hline
\textbf{} &
\multicolumn{2}{c|}{\(\mathrm{NS}_{L/R}\)} &
\multicolumn{2}{c|}{\(\mathrm{R}_{L/R}\)} \\

\(\text{oscillator}\)& \(\text{moding}\) & \(\min\Delta L_{0}\) & \(\text{moding}\) & \(\min\Delta L_{0}\) \\

\hline
\(Z_{a}\) & \(\mathbb{Z}\) & \(1\) & \(\mathbb{Z}\) & \(1\) \\

\((Z_{4},\bar{Z}_5)\) & \(\mathbb{Z}+\tfrac{n}{L}\) & \(1-\frac{n}{L}\) & \(\mathbb{Z}+\tfrac{n}{L}\) & \(1-\frac{n}{L}\) \\

\((Z_{5},\bar{Z}_4)\) & \(\mathbb{Z}-\tfrac{n}{L}\) & \(\frac{n}{L}\) & \(\mathbb{Z}-\tfrac{n}{L}\) & \(\frac{n}{L}\) \\
 
\(\psi^{a}\) & \(\mathbb{Z}+\tfrac{1}{2}\) & \(\tfrac{1}{2}\) & \(\mathbb{Z}\) & \(1\) \\

\((\Psi_{4},\bar{\Psi}_{5})\) & \(\mathbb{Z}+\tfrac{1}{2}+\tfrac{n}{L}\) &
\(\tfrac{1}{2}-\tfrac{n}{L} \)&
\(\mathbb{Z}+\tfrac{n}{L}\) & \(1-\tfrac{n}{L}\) \\

\((\Psi_{5},\bar{\Psi}_{4})\) & \(\mathbb{Z}+\tfrac{1}{2}-\tfrac{n}{L}\) & \(\tfrac{1}{2}+\tfrac{n}{L}\) &
\(\mathbb{Z}-\tfrac{n}{L}\) & \(\frac{n}{L}\) \\
\hline
\end{tabular}
\caption{Modings and mass gap for the different types of excitation modes. This table applies for $\tfrac{n}{L}\leq\tfrac{1}{2}$. For $\tfrac{n}{L}>\tfrac{1}{2}$, the fermionic modes with mode number $-(-\tfrac{1}{2}+\tfrac{n}{L})$ become new lowest-level creation modes and similarly, $-(\tfrac{1}{2}-\tfrac{n}{L})$ becomes an annihilator,  turning the $-(\tfrac{3}{2}-\tfrac{n}{L})$ mode into the lowest-level creation mode.\label{Tab:modings}}
\end{table}

The fractional modes shift the zero-point energy of the (left-moving) NS vacuum in the $n^{th}$ twisted sector by
\begin{equation}\begin{split}
   \delta h_{\rm{NS}}\left(\frac{n}{L}\right)=&\zeta_H\left(-1,\frac{n}{L}\right)+\zeta_{H}\left(-1,1-\frac{n}{L}\right)-2\zeta(-1)\\&\quad+2\zeta_{H}\left(-1,\frac{1}{2}\right)-\zeta_{H}(-1,\alpha_+)-\zeta_H(-1,\alpha_-)\\
   =&\mathrm{min}\left(\frac{n}{L},1-\frac{n}{L}\right)\,,
\end{split}\end{equation}
where $\zeta_H$ is the Hurwitz $\zeta$-function and we have used $\alpha_{+}=\frac{n}{L}+\frac{1}{2},\,\alpha_{-}=\frac{1}{2}-\frac{n}{L}$ for $\frac{n}{L}\leq \frac{1}{2}$ and $\alpha_{+}=\frac{n}{L}-\frac{1}{2},\,\alpha_{-}=\frac{3}{2}-\frac{n}{L}$ for $\frac{n}{L}> \frac{1}{2}$, in line with Table \ref{Tab:modings}.

The standard type IIB GSO projection condition keeps NS states with odd number of worldsheet fermions ensuring that the twisted NS vacuum $\ket{0;k}_{\rm{NS}}^{(n)}$ is projected out. To generate the first states that survive GSO, one thus needs to act with the lowest fermionic oscillators on the twisted vacuum $\ket{0;k}_{\rm{NS}}^{(n)}$. We end up with two massless states in the  left-moving NS sector
\begin{equation}
(\Psi_{4})_{-\left(\tfrac{1}{2}-\tfrac{n}{L}\right)}\ket{0;k}_{\rm{NS}}^{(n)}\,,\qquad (\bar{\Psi}_{5})_{-\left(\tfrac{1}{2}-\tfrac{n}{L}\right)}\ket{0;k}_{\rm{NS}}^{(n)}\,,
\end{equation}
for $\tfrac{n}{L}\leq\tfrac{1}{2}$ and analogous expressions involving $\{\Psi_{5},\bar{\Psi}_{4}\}$ for $\tfrac{n}{L}>\tfrac{1}{2}$.

In the R sector, the shift of the zero-point energy due to the fractional modes ensures that the twisted R vacuum is exactly massless in every twisted sector. However, only two fermionic zero modes arise, resulting in four possible ground state polarisations. The GSO projection imposes a chirality condition, reducing the spinor degeneracy to two states in the left-moving R sector.

At this stage, one therefore has two massless states in each NS/R sector. Accounting for all the twisted sectors and the appropriate tensor products to form closed string states, we find a total of $16(L-1)$ $\mathbb{Z}_{L}$-invariant massless states, which organise into $(L-1)$ tensor multiplets of 6d $\mathcal{N}=(2,0)$ supergravity. 

Having outlined the massless spectrum, we could now generate infinite towers of massive states by acting with the different oscillators on the R/NS ground states. As presented in Table \ref{Tab:modings}, the various towers of excited states start at masses
\begin{equation}
    m^2=\frac{4}{\alpha'} \min\Delta L_{0}\geq \frac{4}{\alpha'} \mathrm{min}\left(\frac{n}{L},1-\frac{n}{L}\right)\,,
\end{equation}
and result in a plethora of energy levels with generic level separation $\Delta m^2=\tfrac{4}{\alpha'L}$. We note, in particular, that the energy spectrum becomes continuous in the limit $L\to \infty$ with fixed $\alpha'$, which will be the topic of Appendix \ref{app: largeL}.

The spectrum of excited states is most conveniently summarised by the torus  partition function
\begin{equation}\label{eq: partition function}
\mathcal{Z}_{\mathcal{H}}(\tau,\bar{\tau})=\mathrm{Tr}_{\mathcal{H}}\left(q^{L_{0}-1}\bar{q}^{\bar{L}_{0}-1}\right)\,,
\end{equation}
where $q=e^{i\pi \tau}$ depends on the torus modulus $\tau$ and $\mathcal{H}$ denotes the Hilbert space of physical states. As discussed above, string theory on a $\mathbb{Z}_L$ orbifold background can be constructed by taking the flat space theory and projecting to $\mathbb{Z}_L$-invariant states, resulting in the untwisted sector. This is captured by inserting the projection operator \eqref{eq:projector} into the partition function \eqref{eq: partition function}, and is nothing else than imposing $\Gamma_L^n$-twisted boundary conditions on the timelike circle and summing over all choices of $n$. In order to guarantee modular invariance, we then similarly have to add $(L-1)$ twisted sectors, where we impose closure up to $\Gamma_L^m$ on the spacelike circle. 

For convenience, we may define $\mathcal{Z}_{\frac{m}{L},\frac{n}{L}}$ as the partition function \eqref{eq: partition function} imposing $\Gamma_L^m$- and $\Gamma_L^n$-twisted boundary conditions on the timelike and spacelike circles, respectively. The $\mathbb{Z}_L$ orbifold partition function then takes the form \cite{Dixon:1985jw,Dixon:1986jc}
\begin{equation}\label{eq: Orbpart}
    \mathcal{Z}_{\text{orb}}=\frac{1}{L}\sum_{m,n=0}^{L-1}\mathcal{Z}_{\frac{m}{L},\frac{n}{L}}\,.
\end{equation}

On can directly generalise the construction of the type IIB partition function $\mathcal{Z}_{0,0}$ to twisted sectors by introducing the elliptic theta function
\begin{equation}
    d_{a,b}(\tfrac{m}{L},\tfrac{n}{L})=\theta\begin{bmatrix}
    \tfrac{a}{2}+\tfrac{m}{L}\\
    \tfrac{b}{2}+\tfrac{n}{L}
    \end{bmatrix}(0\vert\tau)=\sum_{l=-\infty}^\infty q^{(l+\frac{a}{2}+\frac{m}{L})^2}e^{2\pi i (l+\frac{a}{2}+\frac{m}{L})(\frac{b}{2}+\frac{n}{L})}\,.
\end{equation}
The torus partition functions in the individual sectors then takes the explicit form
\begin{equation}\label{eq: partitions}
    \mathcal{Z}_{\frac{m}{L},\frac{n}{L}}=\frac{1}{\vert\mathrm{Im}\tau\vert^{2}}\left|\frac{f(\tfrac{m}{L},\tfrac{n}{L})}{2\eta(\tau)^{6}}\frac{\sum_{a,b}(-1)^{a+b+ab}d_{a,b}(0,0)^{2}d_{a,b}(\tfrac{m}{L},\tfrac{n}{L})d_{a,b}(-\tfrac{m}{L},-\tfrac{n}{L})}{d_{1,1}(\tfrac{m}{L},\tfrac{n}{L})d_{1,1}(-\tfrac{m}{L},-\tfrac{n}{L})+\mathrm{Im}\,\tau\,\eta(\tau)^{6}\delta^{0}_{m}\delta^{0}_{n}}\right|^{2}\,.
\end{equation}
Here, $\eta(\tau)$ denotes the Dedekind $\eta$-function which captures bosonic modes. The sum in the numerator encodes the GSO projection. The factor 
\begin{equation}
    f(\tfrac{m}{L},\tfrac{n}{L})=\begin{cases} 4\sin^2(\tfrac{n}{L}\pi)\,,\quad m=0, n\neq 0\\
        1\,,\qquad \,\,\qquad \text{ else}\,,
    \end{cases}
\end{equation}
captures the contribution of bosonic zero modes, which only appear at $m=0$. We may now explicitly compute the orbifold partition function \eqref{eq: Orbpart} and find that $\mathcal{Z}_{\rm{orb}}=0$. Note that the torus partition function naturally includes antiperiodic boundary conditions for fermions on the time-like circle and therefore is more akin to a supersymmetric index than an unrefined partition function. The vanishing of the partition function is then understood as a direct consequence of the unbroken supersymmetry of the orbifold theory.

To investigate the actual spectrum of states, it is more useful to cancel this sign factor $(-1)^F$ in \eqref{eq: partition function} and consider the unrefined partition function $\tilde{\mathcal{Z}}$, which can be computed by shifting $d_{a,b}\to d_{a,b-1}$ in the numerator of \eqref{eq: partitions}. One can then perform an expansion in small $q$ and determine the energy spectrum from the exponents and the multiplicities from the coefficients. As an example, for $L>4$ a typical untwisted and twisted sector spectrum takes the form \footnote{The orbifolds with $L\leq4$ feature larger degeneracies, essentially due to coincidences within the mode expansion.}
\begin{align}
    \tilde{\mathcal{Z}}_{0}&=\frac{1}{L}\sum_{n=0}^{L-1} \tilde{\mathcal{Z}}_{0,\frac{n}{L}}\sim 96+ 17920 (q\bar{q})^{4} +\dots\,,\\
    \tilde{\mathcal{Z}}_{\frac{1}{L}}&=\frac{1}{L}\sum_{n=0}^{L-1} \tilde{\mathcal{Z}}_{\frac{1}{L},\frac{n}{L}}\sim 16+ 256 (q\bar{q})^{\frac{4}{L}}+1024 (q\bar{q})^{\frac{8}{L}}+\dots\,.
\end{align}
The massless spectrum matches precisely the explicit construction performed above. 
\section{Comments on the large-$L$ limit}\label{app: largeL}
An interesting limit of the $\mathbb{Z}_L$ orbifold theories discussed in this paper is that of 
$L\to \infty$, where the opening angle of the orbifold singularity shrinks to $0$ while the number of twisted sectors grows infinitely. In this limit, the dual gauge theory description is given in terms of ``long'' circular quivers with infinitely many gauge nodes.

In this paper we have discussed the construction of low-energy effective theories describing the massless modes of string theory. The restriction to the massless level is justified by the finite mass gap to the first excited level, which in the orbifold theory on $\mathbb{R}^{1,5}\times \mathbb{C}^2/\mathbb{Z}_L$ is given by
\begin{equation}\label{eq: mass-gap}
    \Delta m^2=\frac{4}{L\alpha'}\,.
\end{equation}
This allows us to choose a cut-off $\Lambda$ below this scale and integrate out all excited states which are heavier than this cut-off. If one were to take $L\to\infty$ at fixed $\alpha'$, the excited spectrum of the string becomes continuous, invalidating this naive low-energy expansion. However, we may take inspiration from the parallel behaviour of zero-modes in a Kaluza-Klein compactification on a circle of radius $R$, which allows for momentum modes with masses $m_{\text{KK}}^2=\tfrac{n^2}{R^2}$ and winding modes with masses $m_{\text{w}}^2=\frac{R^2 w^2}{\alpha'^2}$. As $R\to\infty$, the momentum modes generate a continuum which corresponds to the momentum along the decompactified circle. If $R\to0$, however, the winding modes become continuous, suggesting the appearance of a ``T-dual''  non-compact dimension. A similar phenomenon is now expected to occur in the large-$L$ limit of the orbifold theory, suggesting that the twist number $\tfrac{n}{L}$, enumerating the twisted sectors, may be interpreted as a continuous variable parametrising an ``emergent'' dimension, which replaces the vanishing angular dimension around the orbifold singularity. This ``dual'' geometry is yet to be understood completely.

In the context of AdS/CFT, the large-$L$ limit has been discussed on the gauge theory side, where it is sometimes called a ``deconstruction'' limit of the quiver gauge theory \cite{Hill:2000mu, Arkani-Hamed:2001kyx}, hinting at an effective 5d theory. This limit was also probed by interpolating localisation results like \eqref{eq: Zl-case} to large $L$ \cite{Korchemsky:2025eyc}. Interestingly, in \cite{Korchemsky:2025eyc}, evidence for an emergent fifth dimension was only found in the particular double scaling limit where
\begin{equation}\label{eq: Korchemskyregime}
    \frac{\sqrt{\lambda}}{L}=\frac{R_{\rm{AdS}}}{L\alpha'}\ll 0\,,
\end{equation}
at strong coupling. At fixed $\rm{AdS}$ radius $R_{\rm{AdS}}$, this limit corresponds precisely to the vanishing of the twisted sector mass gap \eqref{eq: mass-gap} $\Delta m^2\to 0$. The emergence of a fifth dimension in gauge theory is thus directly associated to the large-$L$ duality in string theory.

If, on the other hand, one performs a scaling limit with finite $\frac{\sqrt{\lambda}}{L}$, the mass gap \eqref{eq: mass-gap} remains finite and one may take $L\to\infty$ at the level of the effective low-energy theory. Let us outline how our construction of the effective 6d theory in Section \ref{sec: 2} behaves in this limit.

\subsection{Effective action at large $L$}
We now consider the continuous $L\to\infty$ limit on the resolution space and investigate how the field configurations \eqref{eq:GH2normalisedB2field} and dynamics of the twisted scalars are affected. In order to approach this limit, we choose a linear distribution of instantons on $\mathbb{R}^{3}$ parametrised by $\vec{x}_{n}=\big(0,0,\frac{a}{L-1}(2n-L-1)\big)$,\footnote{This preserves $\mathrm{U(1)}\subset \mathrm{SO(3)}$ rotational symmetry of the base, but breaks manifest $\mathbb{Z}_{L}$ symmetry.} where $a$ denotes the resolution parameter. In this configuration, we introduce an ``interpolating'' coordinate 
\begin{equation}
\sigma=\frac{2n-L-1}{L-1}\in[-1,1]\,,
\end{equation}
which becomes continuous in the large-$L$ regime, where the configuration of centers along a line degenerates into a smooth linear distribution of instanton charge of fixed length $2a$ along the $z$-axis.\footnote{The discrete $L\to\infty$ limit of \eqref{GH} was considered in \cite{Anderson}, where the resolved space features an infinite discrete family of compact two-cycles. The metric becomes asymptotically locally flat (ALF) at infinity, approaching $\mathbb{R}^{3}\times S ^{1}$, with the circle at infinity having fixed radius. In our case, the ``smeared'' distribution of centers along the line of charge results in a continuum distribution of two-cycles along the resolved line.} This coordinate labels the position along the line segment at which the sign of the auxiliary potential $\eqref{eq:arbLauxpot}$ changes, playing the role of the discrete twist index $n$ in the continuum limit. We will also assume a finite charge density $\rho=\tfrac{L-1}{4\pi a}\frac{\delta(r)}{r}H(z-a)H(z+a)$ on the linear distribution of instantons. In this regime, the scalar, vector and auxiliary potentials in \eqref{eq:cylindricalcoords} and \eqref{eq:arbLauxpot} become \footnote{These quantities are only approximations in the large-$L$ limit. A proper resummation treatment shows there are subleading $\order{L^0}$ terms which become suppressed and are therefore dropped in this limit.}
\begin{equation}
\label{eq:continuousUandW}
\begin{split}
&U(r,z)=\int\dd^{3}\vec{x}'\frac{\rho}{\sqrt{r^2+(z-z')^2}}=\frac{L-1}{2a}\bigg[\sinh^{-1}\bigg(\frac{z+a}{r}\bigg)-\sinh^{-1}\bigg(\frac{z-a}{r}\bigg)\bigg]\,,\\
&w\cdot\dd x=\frac{L-1}{2a}\bigg[\sqrt{r^{2}+(z+a)^{2}}-\sqrt{r^{2}+(z-a)^{2}}\bigg]\dd\varphi\,,\\
&W_{n}(x)\to W(\sigma;r,z)=\int\dd^{3}\vec{x}'\rho\frac{\mathrm{sgn}(z'-\sigma)}{\sqrt{r^{2}+(z-z')^{2}}}\\&\;\;\,\quad\quad\quad\quad\quad\quad\quad\quad=\frac{L-1}{2a}\bigg[2\sinh^{-1}\bigg(\frac{z-\sigma}{r}\bigg)-\sinh^{-1}\bigg(\frac{z-a}{r}\bigg)-\sinh^{-1}\bigg(\frac{z+a}{r}\bigg)\bigg]\,,
\end{split}
\end{equation}
where we replaced the constants $c_{i,n}$ of \eqref{eq:arbLauxpot} by $\mathrm{sgn}(z'-\sigma)$. One may thus find an explicit anti-self-dual solution to  the vacuum Einstein's equations in the continuous $L\to\infty$ limit of \eqref{GH}, taking the form
\begin{equation}
\label{eq:largeLmetric}
\resizebox{\linewidth}{!}{$
\begin{aligned}
\dd s^{2}=\frac{L-1}{2a}\Bigg[&
\left(\sinh^{-1}\!\left(\frac{z+a}{r}\right)-\sinh^{-1}\!\left(\frac{z-a}{r}\right)\right)^{-1}
\left[\dd\tau'+\left(\sqrt{r^{2}+(z+a)^{2}}-\sqrt{r^{2}+(z-a)^{2}}\right)\dd\varphi\right]^{2}\\
&+\left(\sinh^{-1}\!\left(\frac{z+a}{r}\right)-\sinh^{-1}\!\left(\frac{z-a}{r}\right)\right)
\left(\dd r^{2}+r^{2}\dd\varphi^{2}+\dd z^{2}\right)\Bigg]\,,
\end{aligned}
$}
\end{equation}
where we have re-scaled the fiber coordinate $\tau\to\tau'\in[0,\tfrac{2a}{L-1}4\pi)$ in order to extract a common factor of $\tfrac{L-1}{2a}$ from the metric.
The total $B_{2}$-field becomes a continuous distribution of the individual wrapped components along the line of ``instanton charge'', and admits a representation of the form
\begin{equation}
\label{eq:totalcontB2field}
B_{2}(y,r,z)=\frac{L-1}{2}\int_{-1}^{1}\dd\sigma\,B_{2}(\sigma;y,r,z)\,,
\end{equation}
where its components $B_{2}(\sigma;y,r,z)$ now depend on $\sigma$ and are given as
\begin{equation}
\label{eq:largeLB2cpts}
B_{2}(\sigma;y,r,z)=\pi\alpha'\,b(\sigma;y)\bigg[\mathcal{A}(\sigma;r,z)(e^{1}\wedge e^{2}-e^{3}\wedge e^{4})+\mathcal{B}(\sigma;r,z)(e^{2}\wedge e^{3}-e^{1}\wedge e^{4})\bigg]\,,
\end{equation}
where $\mathcal{A}(\sigma;r,z)=\partial_{z}\Psi(\sigma;r,z)$ and $\mathcal{B}(\sigma;r,z)=\partial_{r}\Psi(\sigma;r,z)$ with
\begin{equation}
\Psi(\sigma;r,z)=\frac{\sinh^{-1}(\tfrac{z-\sigma}{r})-\sinh^{-1}(\tfrac{z-a}{r})}{\sinh^{-1}(\tfrac{z+a}{r})-\sinh^{-1}(\tfrac{z-a}{r})}-\frac{1}{2}\,.
\end{equation}
We choose to re-scale the components as $B_{2}\to(\tfrac{2}{L-1})^{3/2} aB_{2}$ for later convenience. In order to arrive at the large-$L$ analogue of \eqref{eq:discrete6daction}, we are thus interested in computing the integral
\begin{equation}
\int\dd^{10}x\,\vert\dd B_{2}\vert^{2}\to\frac{2a}{L-1}\int_{-1}^{1}\dd\sigma\int_{-1}^{1}\dd\sigma'\int\dd^{10}x\,\dd B_{2}(\sigma;y,r,z)\wedge \star_{10}\,\dd B_{2}(\sigma';y,r,z)\,,
\end{equation}
making use of the field configurations \eqref{eq:largeLB2cpts}. Omitting the $\mathbb{R}^{1,5}$ dependence for the moment, we need to evaluate the large-$L$ analogue of \eqref{eq:GHL2formOverlaps}, which reads
\begin{equation}
\label{eq:largeLintegrand}
\frac{2a}{L-1}\int_{-1}^{1}\dd\sigma\int_{-1}^{1}\dd\sigma'\int_{0}^{\infty}\dd r\int_{-\infty}^{\infty}\dd z\bigg[rU(\mathcal{A}(\sigma)\mathcal{A}(\sigma')+\mathcal{B}(\sigma)\mathcal{B}(\sigma'))\bigg],
\end{equation}
where we have dropped the $(r,z)$ dependence of the various functions. Our choice of re-scaling the components has made the integral \eqref{eq:largeLintegrand} finite in the orbifold limit $a\to0$.

The metric \eqref{eq:largeLmetric} blows up near the line of continuous charge, so one may worry that the integrand of \eqref{eq:largeLintegrand} could be divergent. A local expansion around the singular locus proves the absence of putative divergences. We have evaluated \eqref{eq:largeLintegrand} numerically for arbitrary fixed values of $\sigma,\sigma'\in[-1,1]$, allowing to extract its $\sigma',\sigma$ profiles respectively. 

\begin{figure}
\centering
\begin{subfigure}{.45\textwidth}
  \centering
  \includegraphics[width=.99\linewidth]{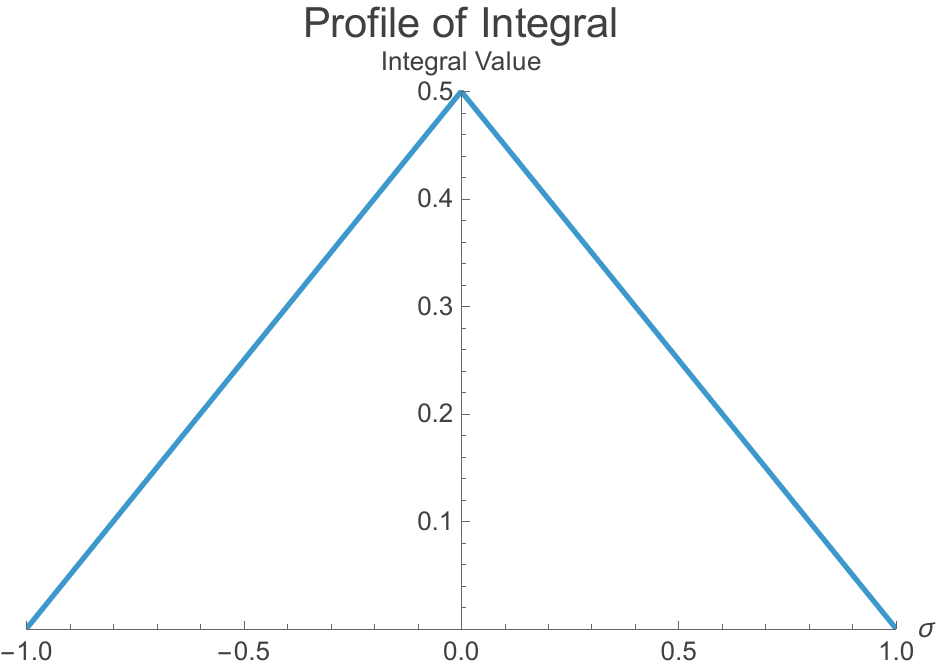}
\end{subfigure}%
\begin{subfigure}{.45\textwidth}
  \centering
  \includegraphics[width=.99\linewidth]{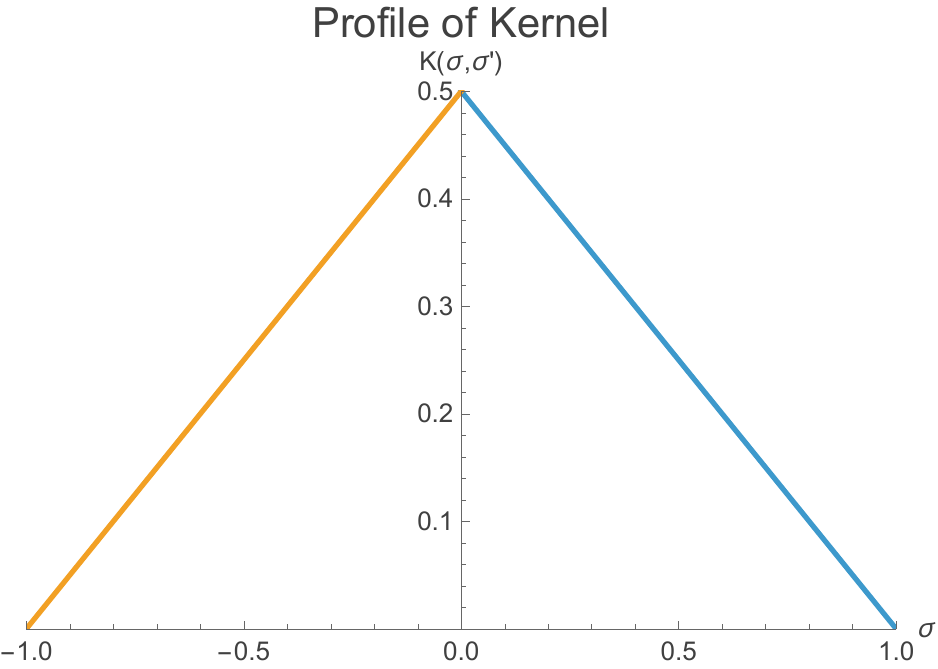}
\end{subfigure}
\caption{Numerical integration (left) showing $\sigma$ profile of \eqref{eq:largeLintegrand} for $a=1$ and $\sigma'=0$. The analytic plot (right) is obtained after transforming the kernel \eqref{eq:continousKernel} according to \eqref{eq:contcoordsrelation} for the same values of $a$ and $\sigma'$. In this case the functions are simply $\tfrac{1+\sigma}{2}$ for $\sigma\in[-1,0]$ and $\tfrac{1-\sigma}{2}$ for $\sigma\in[0,1]$. Remarkably, these match exactly. We have checked this matching for various different values of $\sigma$ and $\sigma'$.}
\label{fig:plots}
\end{figure}

We observe that taking the continuous large-$L$ limit of \eqref{eq:10daction} and then performing the overlap integral \eqref{eq:largeLintegrand} over the $\mathrm{GH}_{L\to\infty}$ resolution leads to an effective action of the form
\begin{equation}
\label{eq:7deffectiveaction}
S_{\mathrm{eff}}\supset \frac{(4\pi^{2}\alpha')^{2}}{2\kappa_{10}^{2}}\int_{-1}^{1}\dd\sigma\int_{-1}^{1}\dd\sigma'\,\mathcal{K}(\sigma,\sigma')\int\dd^{6}y\sqrt{g_{6}}\,\frac{1}{2}\,\nabla b(\sigma;y)\cdot\nabla b(\sigma';y)\,.
\end{equation}
The explicit algebraic form of the kernel $\mathcal{K}(\sigma,\sigma')$ is yet to be determined, but its profile for $\sigma'=0$ and $a=1$ is shown in Figure \ref{fig:plots}.

We now compare this result to the continuous large-$L$ limit at the level of the discrete action \eqref{eq:discrete6daction}. A priori, it is not clear that dimensional reduction and taking $L\to\infty$ should be commuting procedures, but we will nevertheless find agreement in this case. After noting that the twist numbers $\frac{n}{L},\,\frac{m}{L}\in[0,1)$ become continuous in this limit, one finds an action of the form
\begin{equation}\label{eq:cont6daction}
S_{\mathrm{eff}}\supset \frac{1}{2\kappa_{6}^{2}}\int_{0}^{1}\dd \left(\frac{n}{L}\right)\int_{0}^{1}\dd \left(\frac{m}{L}\right)\,\mathcal{K}\left(\frac{n}{L},\frac{m}{L}\right)\int\dd^{6}y\,\sqrt{g_{6}}\nabla b\left(\frac{n}{L};y\right)\cdot\nabla b\left(\frac{m}{L};y\right)\,,
\end{equation}
where the non-local kernel is given by
\begin{equation}
\label{eq:nonlocalkernel}
\mathcal{K}\left(\frac{n}{L},\frac{m}{L}\right)=\bigg[1-\mathrm{max}\left(\frac{n}{L},\frac{m}{L}\right)\bigg]\mathrm{min}\left(\frac{n}{L},\frac{m}{L}\right)\,.
\end{equation}
Its explicit form follows directly from the limiting procedure, and it is reminiscent of some of the expressions found in \cite{PhysRevD.111.046022}. The derivatives in \eqref{eq:cont6daction} act along $\mathbb{R}^{1,5}$ directions $y^{\mu}$ of the background. We will assume a smooth dependence of the modes $b(\tfrac{n}{L};y)$ on the continuous coordinate $\tfrac{n}{L}$, which may be related to the interpolating coordinate $\sigma$ via
\begin{equation}
\label{eq:contcoordsrelation}
\sigma=2\frac{n}{L}-1\,.
\end{equation}
After this transformation, we observe (see e.g. Figure \ref{fig:plots}) that this kernel matches precisely the result of the explicit geometric construction above. This allows us to determine the explicit algebraic form of $\mathcal{K}(\sigma,\sigma')$ in \eqref{eq:7deffectiveaction} as
\begin{equation}
\label{eq:continousKernel}
\mathcal{K}(\sigma,\sigma')=\frac{1}{4}\left(1-\mathrm{max}(\sigma,\sigma')\right)\left(1+\mathrm{min}(\sigma,\sigma')\right)\,.
\end{equation}
We thus conclude that taking the continuum large-$L$ limit and integrating over the resolution space are commuting procedures.

\subsection{Diagonalisation of the kernels}
Starting with \eqref{eq:discrete6daction}, we may express the effective action in the continuous $L\to\infty$ limit in terms of an integral over a single continuous coordinate by considering the discrete Fourier transform
\begin{equation}\label{eq:discreteFT}
b_{n}(y)=\frac{2}{\sqrt{L}}\bigg(b_{0}+\sum_{\alpha=1}^{L-1}e^{\tfrac{i2\pi\alpha}{L} n}b_{\alpha}(y)\bigg)\,,
\end{equation}
of the scalars $b_{n}(y)$ in the discrete $L$ regime, in line with the analysis of \cite{Billo:2022fnb}. Here $\alpha\in[0,L-1]$ is the twisted sector label in momentum space and $b_{0}$ is a constant zero-mode that does not affect the dynamics. This transformation diagonalises $C^{-1}$ and allows us to evaluate one of the two sums in \eqref{eq:discrete6daction}. Going to large $L$ requires defining continuous momenta $p=\tfrac{2\pi\alpha}{L}\in[-\pi,\pi)$ conjugate to positions $\tfrac{n}{L}$. The effective action \eqref{eq:cont6daction} then becomes
\begin{equation}
\label{eq:cont6dactionFT}
S_{\mathrm{eff}}\supset\frac{1}{2\kappa_{6}^{2}}\frac{L}{2\pi}\int\dd^{6}y\sqrt{g_{6}}\int_{-\pi}^{\pi}\mathrm{d}p\,\frac{1}{4\sin^{2}(\tfrac{p}{2})}\nabla b(p;y)\nabla b^{*}(-p;y)\,,
\end{equation}
where the factor $(4\sin^{2}(\tfrac{p}{2}))^{-1}$ is due to the eigenvalues of $C^{-1}$. This is the expected form of the effective action \eqref{eq:discrete6daction} in the continuous large-$L$ limit in momentum-space using the algebraic geometry language of the flat space resolutions (see e.g. \cite{Korchemsky:2025eyc}). 

On the other hand, we can also introduce an appropriate basis that diagonalises the kernel \eqref{eq:continousKernel}, which was obtained directly from our field configurations \eqref{eq:largeLB2cpts}. This procedure is analogous in spirit to \eqref{eq:discreteFT}. An orthogonal basis of eigenfunctions achieving this diagonalisation takes the form $\sin(\tfrac{\pi n}{2}(\sigma+1))$, so that we can identify
\begin{equation}
\mathcal{K}(\sigma,\sigma')=2\sum_{n=1}^\infty\frac{\sin(\frac{\pi n}{2}(\sigma+1))\sin(\frac{\pi n}{2}(\sigma'+1))}{n^2\pi^2}\,.
\end{equation}
We may thus parametrise the twisted scalars $b(\sigma;y)$ as
\begin{equation}
b(\sigma;y)=\sum_{n=1}^{\infty}\frac{\sqrt{2}}{n\pi}b_{n}(y)\sin(\frac{\pi n}{2}(\sigma+1))\,,
\end{equation}
corresponding to a Fourier series expansion with vanishing Dirichlet boundary conditions. Note that at the extremal values $(\sigma=\pm 1)$ of the interpolating coordinate, the field configuration vanishes. Indeed, one may identify the kernel \eqref{eq:continousKernel} as the Green's function of a one-dimensional Laplacian on a compact interval $\sigma\in[-1,1]$. The effective action \eqref{eq:7deffectiveaction} then takes the form
\begin{equation}
\label{eq:largeLeffectiveaction}
S_{\mathrm{eff}}\supset \frac{1}{2\kappa_{6}^{2}}\int\dd^{6}y\sqrt{g_{6}}\sum_{n=1}^{\infty}\frac{1}{2}\nabla b_{n}(y)\cdot\nabla b_{n}(y)\,,
\end{equation}
featuring an infinite tower of massless fields $b_{n}(y)$ with canonically normalised kinetic terms. This action differs from that in \eqref{eq:cont6dactionFT} by a mere  normalisation $(4\sin^{2}\left(p/2\right))^{-1}$. In locally flat space, there is no obstruction to choosing a canonical normalisation for the $b_n(y)$ field as in \eqref{eq:largeLeffectiveaction}. This situation may change once interaction terms are introduced. In particular, in the resolution of $\mathrm{AdS}_{5}\cross S ^{5}/\mathbb{Z}_{L}$, the factor $(4\sin^{2}\left(p/2\right))^{-1}$ seems crucial to match localisation results from the string side (see \cite{Billo:2022fnb} for finite $L$ and \cite{Korchemsky:2025eyc} for the large-$L$ limit). One may wonder if that factor in the localisation results should be equally attributed to mere convention or whether some physical interpretation of the $(4\sin^{2}\left(p/2\right))^{-1}$ factor is possible.

Having demonstrated how to construct the low-energy effective action for twisted sector scalar fields in the resolved large-$L$ orbifold background, we would like to close by commenting on the inclusion of $\alpha'$-corrections in this case, in order to make contact with Figure \ref{fig:typeIIB_limits}. As discussed above, the main difficulty in analysing the large-$L$ limit is the parametrically shrinking mass-gap \eqref{eq: mass-gap} of the string spectrum. In order to investigate $\alpha'$-corrections, one would have to consider the string worldsheet theory at small (with respect to the AdS scale $R_{\mathrm{AdS}}$) but finite value of $\alpha'$. If we were to take the strict $L\to\infty$ limit, we would immediately enter the regime \eqref{eq: Korchemskyregime}, where our 6d low-energy action breaks down. This matches the observation that at large but finite $L$, we expect $\alpha'$-corrections starting with terms proportional to $\psi^{(2)}\left(\frac{n}{L}\right)+\psi^{(2)}\left(1-\frac{n}{L}\right)$, which diverge for $L\to\infty$ at any finite $n$ \cite{Korchemsky:2025eyc}. Understanding the dual frame in the regime \eqref{eq: Korchemskyregime} thus remains the crucial challenge. 

\bibliographystyle{JHEP}
\bibliography{ZLOrbifoldRefs}

@article{Anderson, title={Complete Ricci-flat Kähler manifolds of infinite topological type.}, volume={1989},
journal={Commun.Math. Phys. 125, 637–642},
DOI={https://doi.org/10.1007/BF01228345},
author={Anderson, M.T. and Kronheimer, P.B. and LeBrun, C.}}

@article{Korchemsky:2025eyc,
    author = "Korchemsky, Gregory P. and Testa, Alessandro",
    title = "{Correlation functions in four-dimensional superconformal long circular quivers}",
    eprint = "2501.17223",
    archivePrefix = "arXiv",
    primaryClass = "hep-th",
    doi = "10.1007/JHEP07(2025)223",
    journal = "JHEP",
    volume = "07",
    pages = "223",
    year = "2025"
}

@article{Maldacena_1999,
    author = "Maldacena, Juan Martin",
    title = "{The Large $N$ limit of superconformal field theories and supergravity}",
    eprint = "hep-th/9711200",
    archivePrefix = "arXiv",
    reportNumber = "HUTP-97-A097, HUTP-98-A097",
    doi = "10.4310/ATMP.1998.v2.n2.a1",
    journal = "Adv. Theor. Math. Phys.",
    volume = "2",
    pages = "231--252",
    year = "1998"
}

@article{Dixon:1986jc,
    author = "Dixon, Lance J. and Harvey, Jeffrey A. and Vafa, C. and Witten, Edward",
    title = "{Strings on Orbifolds. 2.}",
    reportNumber = "PRINT-86-0246 (PRINCETON)",
    doi = "10.1016/0550-3213(86)90287-7",
    journal = "Nucl. Phys. B",
    volume = "274",
    pages = "285--314",
    year = "1986"
}

@article{McKay,
author= "John McKay",
title="
Graphs, singularities, and finite groups.",
volume= "37",
journal ="
Proc. Symp. Pure Math. No. 183.",
year ="1980"
}

@book{Blumenhagen:2013fgp,
    author = {Blumenhagen, Ralph and L{\"u}st, Dieter and Theisen, Stefan},
    title = "{Basic concepts of string theory}",
    doi = "10.1007/978-3-642-29497-6",
    isbn = "978-3-642-29496-9",
    publisher = "Springer",
    address = "Heidelberg, Germany",
    series = "Theoretical and Mathematical Physics",
    year = "2013"
}

@article{Dixon:1986qv,
    author = "Dixon, Lance J. and Friedan, Daniel and Martinec, Emil J. and Shenker, Stephen H.",
    title = "{The Conformal Field Theory of Orbifolds}",
    reportNumber = "EFI-86-42-CHICAGO",
    doi = "10.1016/0550-3213(87)90676-6",
    journal = "Nucl. Phys. B",
    volume = "282",
    pages = "13--73",
    year = "1987"
}

@article{Hamidi:1986vh,
    author = "Hamidi, Shahram and Vafa, Cumrun",
    title = "{Interactions on Orbifolds}",
    reportNumber = "HUTP-86-A041, CALT-68-1349",
    doi = "10.1016/0550-3213(87)90006-X",
    journal = "Nucl. Phys. B",
    volume = "279",
    pages = "465--513",
    year = "1987"
}

@article{Lunin:2000yv,
    author = "Lunin, Oleg and Mathur, Samir D.",
    title = "{Correlation functions for M**N / S(N) orbifolds}",
    eprint = "hep-th/0006196",
    archivePrefix = "arXiv",
    reportNumber = "OHSTPY-HEP-T-00-010",
    doi = "10.1007/s002200100431",
    journal = "Commun. Math. Phys.",
    volume = "219",
    pages = "399--442",
    year = "2001"
}

@article{Skrzypek:2023fkr,
    author = "Skrzypek, Torben and Tseytlin, Arkady A.",
    title = "{On AdS/CFT duality in the twisted sector of string theory on AdS$_{5}${\texttimes} S$^{5}$/{\ensuremath{\mathbb{Z}}}$_{2}$ orbifold background}",
    eprint = "2312.13850",
    archivePrefix = "arXiv",
    primaryClass = "hep-th",
    reportNumber = "Imperial-TP-AT-2023-07",
    doi = "10.1007/JHEP03(2024)045",
    journal = "JHEP",
    volume = "03",
    pages = "045",
    year = "2024"
}

@article{Green:1982sw,
    author = "Green, Michael B. and Schwarz, John H. and Brink, Lars",
    title = "{N=4 Yang-Mills and N=8 Supergravity as Limits of String Theories}",
    reportNumber = "CALT-68-880",
    doi = "10.1016/0550-3213(82)90336-4",
    journal = "Nucl. Phys. B",
    volume = "198",
    pages = "474--492",
    year = "1982"
}

@article{Sakai:1986bi,
    author = "Sakai, N. and Tanii, Y.",
    title = "{One Loop Amplitudes and Effective Action in Superstring Theories}",
    reportNumber = "TIT-HEP-103",
    doi = "10.1016/0550-3213(87)90114-3",
    journal = "Nucl. Phys. B",
    volume = "287",
    pages = "457",
    year = "1987"
}

@article{Billo:2022lrv,
    author = "Billo, M. and Frau, M. and Lerda, A. and Pini, A. and Vallarino, P.",
    title = "{Strong coupling expansions in $ \mathcal{N} $ = 2 quiver gauge theories}",
    eprint = "2211.11795",
    archivePrefix = "arXiv",
    primaryClass = "hep-th",
    doi = "10.1007/JHEP01(2023)119",
    journal = "JHEP",
    volume = "01",
    pages = "119",
    year = "2023"
}

@Article{Skrzypek:2022cgg,
  author        = {Skrzypek, Torben},
  journal       = {J. Phys. A},
  title         = {{Integrability treatment of AdS/CFT orbifolds}},
  year          = {2023},
  number        = {34},
  pages         = {345401},
  volume        = {56},
  archiveprefix = {arXiv},
  doi           = {10.1088/1751-8121/ace947},
  eprint        = {2211.03806},
  primaryclass  = {hep-th},
}

@Article{Liu:2022bfg,
  author        = {Liu, James T. and Minasian, Ruben and Savelli, Raffaele and Schachner, Andreas},
  journal       = {JHEP},
  title         = {{Type IIB at eight derivatives: insights from Superstrings, Superfields and Superparticles}},
  year          = {2022},
  pages         = {267},
  volume        = {08},
  archiveprefix = {arXiv},
  doi           = {10.1007/JHEP08(2022)267},
  eprint        = {2205.11530},
  primaryclass  = {hep-th},
}

@Article{Freeman:1986zh,
  author       = {Freeman, M. D. and Pope, C. N. and Sohnius, M. F. and Stelle, K. S.},
  journal      = {Phys. Lett. B},
  title        = {{Higher Order $\sigma$ Model Counterterms and the Effective Action for Superstrings}},
  year         = {1986},
  pages        = {199--204},
  volume       = {178},
  doi          = {10.1016/0370-2693(86)91495-4},
  reportnumber = {IMPERIAL/TP/85-86/27},
}

@Article{Grisaru:1986vi,
  author       = {Grisaru, Marcus T. and Zanon, D.},
  journal      = {Phys. Lett. B},
  title        = {{$\sigma$ Model Superstring Corrections to the Einstein-hilbert Action}},
  year         = {1986},
  pages        = {347--351},
  volume       = {177},
  doi          = {10.1016/0370-2693(86)90765-3},
  reportnumber = {HUTP-86/A046, BRX-TH-202},
}

@Article{Liu:2013dna,
  author        = {Liu, James T. and Minasian, Ruben},
  journal       = {Nucl. Phys. B},
  title         = {{Higher-derivative couplings in string theory: dualities and the B-field}},
  year          = {2013},
  pages         = {413--470},
  volume        = {874},
  archiveprefix = {arXiv},
  doi           = {10.1016/j.nuclphysb.2013.06.002},
  eprint        = {1304.3137},
  primaryclass  = {hep-th},
  reportnumber  = {IPHT-T13-043-MCTP-13-09},
}

@Article{Liu:2019ses,
  author        = {Liu, James T. and Minasian, Ruben},
  journal       = {Nucl. Phys. B},
  title         = {{Higher-derivative couplings in string theory: five-point contact terms}},
  year          = {2021},
  pages         = {115386},
  volume        = {967},
  archiveprefix = {arXiv},
  doi           = {10.1016/j.nuclphysb.2021.115386},
  eprint        = {1912.10974},
  primaryclass  = {hep-th},
  reportnumber  = {IPhT-t19/168, LCTP-19-36},
}

@Article{Billo:2022gmq,
  author        = {Bill\`o, Marco and Frau, Marialuisa and Lerda, Alberto and Pini, Alessandro and Vallarino, Paolo},
  journal       = {Phys. Rev. Lett.},
  title         = {{Structure Constants in N=2 Superconformal Quiver Theories at Strong Coupling and Holography}},
  year          = {2022},
  number        = {3},
  pages         = {031602},
  volume        = {129},
  archiveprefix = {arXiv},
  doi           = {10.1103/PhysRevLett.129.031602},
  eprint        = {2206.13582},
  primaryclass  = {hep-th},
}

@Article{Beccaria:2021hvt,
  author        = {Beccaria, M. and Bill\`o, M. and Frau, M. and Lerda, A. and Pini, A.},
  journal       = {JHEP},
  title         = {{Exact results in a $ \mathcal{N} $ = 2 superconformal gauge theory at strong coupling}},
  year          = {2021},
  pages         = {185},
  volume        = {07},
  archiveprefix = {arXiv},
  doi           = {10.1007/JHEP07(2021)185},
  eprint        = {2105.15113},
  primaryclass  = {hep-th},
}

@Article{Berenstein:2002jq,
  author        = {Berenstein, David Eliecer and Maldacena, Juan Martin and Nastase, Horatiu Stefan},
  journal       = {JHEP},
  title         = {{Strings in flat space and pp waves from N=4 superYang-Mills}},
  year          = {2002},
  pages         = {013},
  volume        = {04},
  archiveprefix = {arXiv},
  doi           = {10.1088/1126-6708/2002/04/013},
  eprint        = {hep-th/0202021},
}

@Article{Prasad:1979kg,
  author  = {Prasad, M. K.},
  journal = {Phys. Lett. B},
  title   = {{Equivalence of Eguchi-Hanson metric to two-center Gibbons-Hawking metric}},
  year    = {1979},
  pages   = {310--310},
  volume  = {83},
  doi     = {10.1016/0370-2693(79)91114-6},
}

@article{Gibbons:1978tef,
    author = "Gibbons, G. W. and Hawking, S. W.",
    title = "{Gravitational Multi - Instantons}",
    reportNumber = "Print-79-0042 (CAMBRIDGE)",
    doi = "10.1016/0370-2693(78)90478-1",
    journal = "Phys. Lett. B",
    volume = "78",
    pages = "430",
    year = "1978"
}

@article{Eguchi:1978xp,
    author = "Eguchi, Tohru and Hanson, Andrew J.",
    title = "{Asymptotically Flat Selfdual Solutions to Euclidean Gravity}",
    reportNumber = "SLAC-PUB-2087, LBL-7273",
    doi = "10.1016/0370-2693(78)90566-X",
    journal = "Phys. Lett. B",
    volume = "74",
    pages = "249--251",
    year = "1978"
}

@article{Floratos:2002uh,
    author = "Floratos, Emmanuel and Kehagias, Alex",
    title = "{Penrose limits of orbifolds and orientifolds}",
    eprint = "hep-th/0203134",
    archivePrefix = "arXiv",
    doi = "10.1088/1126-6708/2002/07/031",
    journal = "JHEP",
    volume = "07",
    pages = "031",
    year = "2002"
}

@article{Blau:2001ne,
    author = "Blau, Matthias and Figueroa-O'Farrill, Jose M. and Hull, Christopher and Papadopoulos, George",
    title = "{A New maximally supersymmetric background of IIB superstring theory}",
    eprint = "hep-th/0110242",
    archivePrefix = "arXiv",
    reportNumber = "EMPG-01-18, QMUL-PH-01-12",
    doi = "10.1088/1126-6708/2002/01/047",
    journal = "JHEP",
    volume = "01",
    pages = "047",
    year = "2002"
}

@article{Itzhaki:2002kh,
    author = "Itzhaki, N. and Klebanov, Igor R. and Mukhi, Sunil",
    title = "{PP wave limit and enhanced supersymmetry in gauge theories}",
    eprint = "hep-th/0202153",
    archivePrefix = "arXiv",
    reportNumber = "PUPT-2024, TIFR-TH-02-06",
    doi = "10.1088/1126-6708/2002/03/048",
    journal = "JHEP",
    volume = "03",
    pages = "048",
    year = "2002"
}

@Article{Alishahiha:2002ev,
  author        = {Alishahiha, Mohsen and Sheikh-Jabbari, Mohammad M.},
  journal       = {Phys. Lett. B},
  title         = {{The pp wave limits of orbifolded AdS(5)$\times$ S5}},
  year          = {2002},
  pages         = {328--336},
  volume        = {535},
  archiveprefix = {arXiv},
  doi           = {10.1016/S0370-2693(02)01771-9},
  eprint        = {hep-th/0203018},
  reportnumber  = {IPM-P-2002-004, SU-ITP-02-10},
}

@Article{Kim:2002fp,
  author        = {Kim, Nakwoo and Pankiewicz, Ari and Rey, Soo-Jong and Theisen, Stefan},
  journal       = {Eur. Phys. J. C},
  title         = {{Superstring on PP wave orbifold from large N quiver gauge theory}},
  year          = {2002},
  pages         = {327--332},
  volume        = {25},
  archiveprefix = {arXiv},
  doi           = {10.1007/s10052-002-0986-y},
  eprint        = {hep-th/0203080},
  reportnumber  = {AEI-2002-016, SNUST-02-03-01},
}

@Article{Sahraoui:2002sp,
  author        = {Sahraoui, El Mostapha and Saidi, El Hassan},
  journal       = {Phys. Lett. B},
  title         = {{Metrics building of pp wave orbifold geometries}},
  year          = {2003},
  pages         = {221--228},
  volume        = {558},
  archiveprefix = {arXiv},
  doi           = {10.1016/S0370-2693(03)00279-X},
  eprint        = {hep-th/0210168},
  reportnumber  = {UFRHEP-02-05},
}

@Article{Gukov:1998kk,
  author        = {Gukov, Sergei},
  journal       = {Phys. Lett. B},
  title         = {{Comments on N=2 AdS orbifolds}},
  year          = {1998},
  pages         = {23--28},
  volume        = {439},
  archiveprefix = {arXiv},
  doi           = {10.1016/S0370-2693(98)01005-3},
  eprint        = {hep-th/9806180},
  reportnumber  = {PUPT-1800, ITEP-TH-33-98, LANDAU-98-TMP-2},
}

@Article{Kachru:1998ys,
  author        = {Kachru, Shamit and Silverstein, Eva},
  journal       = {Phys. Rev. Lett.},
  title         = {{4-D conformal theories and strings on orbifolds}},
  year          = {1998},
  pages         = {4855--4858},
  volume        = {80},
  archiveprefix = {arXiv},
  doi           = {10.1103/PhysRevLett.80.4855},
  eprint        = {hep-th/9802183},
  reportnumber  = {SLAC-PUB-7756, LBL-41440, LBNL-41440, UCB-PTH-98-12},
}

@Article{Lawrence:1998ja,
  author        = {Lawrence, Albion E. and Nekrasov, Nikita and Vafa, Cumrun},
  journal       = {Nucl. Phys. B},
  title         = {{On conformal field theories in four-dimensions}},
  year          = {1998},
  pages         = {199--209},
  volume        = {533},
  archiveprefix = {arXiv},
  doi           = {10.1016/S0550-3213(98)00495-7},
  eprint        = {hep-th/9803015},
  reportnumber  = {HUTP-98-A015, ITEP-TH-15-98},
}

@article{Bershadsky:1998mb,
    author = "Bershadsky, Michael and Kakushadze, Zurab and Vafa, Cumrun",
    title = "{String expansion as large N expansion of gauge theories}",
    eprint = "hep-th/9803076",
    archivePrefix = "arXiv",
    reportNumber = "HUTP-98-A017, NUB-3170",
    doi = "10.1016/S0550-3213(98)00272-7",
    journal = "Nucl. Phys. B",
    volume = "523",
    pages = "59--72",
    year = "1998"
}

@Article{Bershadsky:1998cb,
  author        = {Bershadsky, Michael and Johansen, Andrei},
  journal       = {Nucl. Phys. B},
  title         = {{Large N limit of orbifold field theories}},
  year          = {1998},
  pages         = {141--148},
  volume        = {536},
  archiveprefix = {arXiv},
  doi           = {10.1016/S0550-3213(98)00526-4},
  eprint        = {hep-th/9803249},
  reportnumber  = {HUTP-98-A038},
}

@Article{Klebanov:1998hh,
  author        = {Klebanov, Igor R. and Witten, Edward},
  journal       = {Nucl. Phys. B},
  title         = {{Superconformal field theory on three-branes at a Calabi-Yau singularity}},
  year          = {1998},
  pages         = {199--218},
  volume        = {536},
  archiveprefix = {arXiv},
  doi           = {10.1016/S0550-3213(98)00654-3},
  eprint        = {hep-th/9807080},
  reportnumber  = {IASSNS-HEP-98-64, PUPT-1804},
}

@Article{Klebanov:1999rd,
  author        = {Klebanov, Igor R. and Nekrasov, Nikita A.},
  journal       = {Nucl. Phys. B},
  title         = {{Gravity duals of fractional branes and logarithmic RG flow}},
  year          = {2000},
  pages         = {263--274},
  volume        = {574},
  archiveprefix = {arXiv},
  doi           = {10.1016/S0550-3213(00)00016-X},
  eprint        = {hep-th/9911096},
  reportnumber  = {PUPT-1897, ITEP-TH-61-99},
}

@Article{Beccaria:2022ypy,
  author        = {Beccaria, M. and Korchemsky, G. P. and Tseytlin, A. A.},
  journal       = {JHEP},
  title         = {{Strong coupling expansion in \ensuremath{\mathscr{N}} = 2 superconformal theories and the Bessel kernel}},
  year          = {2022},
  pages         = {226},
  volume        = {09},
  archiveprefix = {arXiv},
  doi           = {10.1007/JHEP09(2022)226},
  eprint        = {2207.11475},
  primaryclass  = {hep-th},
  reportnumber  = {IPhT--T22/040, Imperial-TP-AT-2022-04},
}

@Article{Beccaria:2023kbl,
  author        = {Beccaria, M. and Korchemsky, G. P. and Tseytlin, A. A.},
  journal       = {JHEP},
  title         = {{Non-planar corrections in orbifold/orientifold $ \mathcal{N} $ = 2 superconformal theories from localization}},
  year          = {2023},
  pages         = {165},
  volume        = {05},
  archiveprefix = {arXiv},
  doi           = {10.1007/JHEP05(2023)165},
  eprint        = {2303.16305},
  primaryclass  = {hep-th},
  reportnumber  = {IPhT-T23/015, Imperial-TP-AT-2023-01},
}

@Article{Galvagno:2020cgq,
  author        = {Galvagno, Francesco and Preti, Michelangelo},
  journal       = {JHEP},
  title         = {{Chiral correlators in $ \mathcal{N} $ = 2 superconformal quivers}},
  year          = {2021},
  pages         = {201},
  volume        = {05},
  archiveprefix = {arXiv},
  doi           = {10.1007/JHEP05(2021)201},
  eprint        = {2012.15792},
  primaryclass  = {hep-th},
  reportnumber  = {NORDITA 2020-080},
}

@Article{Billo:2021rdb,
  author        = {Billo, M. and Frau, M. and Galvagno, F. and Lerda, A. and Pini, A.},
  journal       = {JHEP},
  title         = {{Strong-coupling results for $ \mathcal{N} $ = 2 superconformal quivers and holography}},
  year          = {2021},
  pages         = {161},
  volume        = {10},
  archiveprefix = {arXiv},
  doi           = {10.1007/JHEP10(2021)161},
  eprint        = {2109.00559},
  primaryclass  = {hep-th},
}

@Article{Billo:2022fnb,
  author        = {Billo, M. and Frau, M. and Lerda, A. and Pini, A. and Vallarino, P.},
  journal       = {JHEP},
  title         = {{Localization vs holography in 4d  $ \mathcal{N} $ = 2 quiver theories}},
  year          = {2022},
  pages         = {020},
  volume        = {10},
  archiveprefix = {arXiv},
  doi           = {10.1007/JHEP10(2022)020},
  eprint        = {2207.08846},
  primaryclass  = {hep-th},
}

@article{PhysRevD.111.046022,
  title = {Continuous quiver gauge theories},
  author = {Sobko, Evgeny},
  journal = {Phys. Rev. D},
  volume = {111},
  issue = {4},
  pages = {046022},
  numpages = {6},
  year = {2025},
  month = {Feb},
  publisher = {American Physical Society},
  doi = {10.1103/PhysRevD.111.046022},
  url = {https://link.aps.org/doi/10.1103/PhysRevD.111.046022}
}

@article{Dixon:1985jw,
    author = "Dixon, Lance J. and Harvey, Jeffrey A. and Vafa, C. and Witten, Edward",
    editor = "Schellekens, B.",
    title = "{Strings on Orbifolds}",
    reportNumber = "PRINT-85-0616 (PRINCETON)",
    doi = "10.1016/0550-3213(85)90593-0",
    journal = "Nucl. Phys. B",
    volume = "261",
    pages = "678--686",
    year = "1985"
}

@Article{Douglas:1996sw,
  author        = {Douglas, Michael R. and Moore, Gregory W.},
  title         = {{D-branes, quivers, and ALE instantons}},
  year          = {1996},
  month         = {3},
  archiveprefix = {arXiv},
  eprint        = {hep-th/9603167},
  reportnumber  = {RU-96-15, YCTP-P5-96},
}

@Article{Gross:1986iv,
  author       = {Gross, David J. and Witten, Edward},
  journal      = {Nucl. Phys. B},
  title        = {{Superstring Modifications of Einstein's Equations}},
  year         = {1986},
  pages        = {1},
  volume       = {277},
  doi          = {10.1016/0550-3213(86)90429-3},
  reportnumber = {Print-86-0250 (PRINCETON)},
}

@article{Witten:1998qj,
    author = "Witten, Edward",
    title = "{Anti de Sitter space and holography}",
    eprint = "hep-th/9802150",
    archivePrefix = "arXiv",
    reportNumber = "IASSNS-HEP-98-15",
    doi = "10.4310/ATMP.1998.v2.n2.a2",
    journal = "Adv. Theor. Math. Phys.",
    volume = "2",
    pages = "253--291",
    year = "1998"
}

@article{Eguchi:1978gw,
    author = "Eguchi, Tohru and Hanson, Andrew J.",
    title = "{Selfdual Solutions to Euclidean Gravity}",
    reportNumber = "SLAC-PUB-2213, LBL-8265",
    doi = "10.1016/0003-4916(79)90282-3",
    journal = "Annals Phys.",
    volume = "120",
    pages = "82",
    year = "1979"
}

@article{Gadde:2009dj,
    author = "Gadde, Abhijit and Pomoni, Elli and Rastelli, Leonardo",
    title = "{The Veneziano Limit of N = 2 Superconformal QCD: Towards the String Dual of N = 2 SU(N(c)) SYM with N(f) = 2 N(c)}",
    eprint = "0912.4918",
    archivePrefix = "arXiv",
    primaryClass = "hep-th",
    reportNumber = "YITP-SB-09-48",
    month = "12",
    year = "2009"
}

@article{Gadde:2010zi,
    author = "Gadde, Abhijit and Pomoni, Elli and Rastelli, Leonardo",
    title = "{Spin Chains in $\mathcal{N}$=2 Superconformal Theories: From the $\mathbb{Z}_{2}$ Quiver to Superconformal QCD}",
    eprint = "1006.0015",
    archivePrefix = "arXiv",
    primaryClass = "hep-th",
    reportNumber = "YITP-SB-10-20",
    doi = "10.1007/JHEP06(2012)107",
    journal = "JHEP",
    volume = "06",
    pages = "107",
    year = "2012"
}

@article{Pomoni:2021pbj,
    author = "Pomoni, Elli and Rabe, Randle and Zoubos, Konstantinos",
    title = "{Dynamical spin chains in 4D $ \mathcal{N} $ = 2 SCFTs}",
    eprint = "2106.08449",
    archivePrefix = "arXiv",
    primaryClass = "hep-th",
    reportNumber = "DESY-21-082, DESY 21-082",
    doi = "10.1007/JHEP08(2021)127",
    journal = "JHEP",
    volume = "08",
    pages = "127",
    year = "2021"
}

@article{Bertle:2024djm,
    author = "Bertle, Hanno and Pomoni, Elli and Zhang, Xinyu and Zoubos, Konstantinos",
    title = "{Hidden symmetries of 4D $ \mathcal{N} $ = 2 gauge theories}",
    eprint = "2411.11612",
    archivePrefix = "arXiv",
    primaryClass = "hep-th",
    reportNumber = "DESY-24-172",
    doi = "10.1007/JHEP02(2025)205",
    journal = "JHEP",
    volume = "02",
    pages = "205",
    year = "2025"
}

@article{Bozkurt:2024tpz,
    author = "Bozkurt, Deniz N. and Nieto Garc{\'\i}a, Juan Miguel and Pomoni, Elli",
    title = "{Long-range to the Rescue of Yang-Baxter}",
    eprint = "2408.03365",
    archivePrefix = "arXiv",
    primaryClass = "hep-th",
    reportNumber = "DESY-24-113, ZMP-HH/24-16",
    month = "8",
    year = "2024"
}

@article{Bozkurt:2025exl,
    author = "Bozkurt, Deniz N. and Nieto Garc{\'\i}a, Juan Miguel and Kong, Ziwen and Pomoni, Elli",
    title = "{Long-range to the Rescue of Yang-Baxter II}",
    eprint = "2507.08934",
    archivePrefix = "arXiv",
    primaryClass = "hep-th",
    reportNumber = "DESY 25-083, ZMP-HH/25-10",
    month = "7",
    year = "2025"
}

@article{lePlat:2025eod,
    author = "le Plat, Dennis and Skrzypek, Torben",
    title = "{Three-point functions from integrability in $\mathcal{N}=2$ orbifold theories}",
    eprint = "2506.21323",
    archivePrefix = "arXiv",
    primaryClass = "hep-th",
    reportNumber = "DESY-25-088",
    month = "6",
    year = "2025"
}

@article{Ferrando:2025qkr,
    author = {Ferrando, Gwena{\"e}l and Komatsu, Shota and Lefundes, Gabriel and Serban, Didina},
    title = "{Exact Three-Point Functions in $\mathcal{N}=2$ Superconformal Field Theories: Integrability vs. Localization}",
    eprint = "2503.07295",
    archivePrefix = "arXiv",
    primaryClass = "hep-th",
    reportNumber = "BONN-TH-2025-09",
    month = "3",
    year = "2025"
}

@article{Hill:2000mu,
    author = "Hill, Christopher T. and Pokorski, Stefan and Wang, Jing",
    title = "{Gauge Invariant Effective Lagrangian for Kaluza-Klein Modes}",
    eprint = "hep-th/0104035",
    archivePrefix = "arXiv",
    reportNumber = "FERMILAB-PUB-01-043-T",
    doi = "10.1103/PhysRevD.64.105005",
    journal = "Phys. Rev. D",
    volume = "64",
    pages = "105005",
    year = "2001"
}

@article{Arkani-Hamed:2001kyx,
    author = "Arkani-Hamed, Nima and Cohen, Andrew G. and Georgi, Howard",
    title = "{(De)constructing dimensions}",
    eprint = "hep-th/0104005",
    archivePrefix = "arXiv",
    reportNumber = "HUTP-01-A015, BUHEP-01-05, LBNL-47676, UCB-PTH-01-11",
    doi = "10.1103/PhysRevLett.86.4757",
    journal = "Phys. Rev. Lett.",
    volume = "86",
    pages = "4757--4761",
    year = "2001"
}

@article{Alday:2023mvu,
    author = "Alday, Luis F. and Hansen, Tobias",
    title = "{The AdS Virasoro-Shapiro amplitude}",
    eprint = "2306.12786",
    archivePrefix = "arXiv",
    primaryClass = "hep-th",
    doi = "10.1007/JHEP10(2023)023",
    journal = "JHEP",
    volume = "10",
    pages = "023",
    year = "2023"
}

@article{Zarembo:2020tpf,
    author = "Zarembo, K.",
    title = "{Quiver CFT at strong coupling}",
    eprint = "2003.00993",
    archivePrefix = "arXiv",
    primaryClass = "hep-th",
    reportNumber = "NORDITA 2020-022",
    doi = "10.1007/JHEP06(2020)055",
    journal = "JHEP",
    volume = "06",
    pages = "055",
    year = "2020"
}
\end{document}